\documentclass[11pt]{article}
\pdfoutput=1 
\usepackage{jheppub} 
\usepackage{graphicx}
\usepackage{enumerate}
\usepackage{hyperref}
\usepackage{amsmath}
\usepackage{amssymb}
\usepackage{xcolor}

\def\beq{\begin{equation}\begin{aligned}}
\def\eeq{\end{aligned}\end{equation}}
\def\barray{\begin{eqnarray}}
\def\earray{\end{eqnarray}}
\def\barrayn{\begin{eqnarray*}}
\def\earrayn{\end{eqnarray*}}

\newcommand{\MeV}{\text{MeV}}

\newcommand{\MPl}{M_{\rm Pl}}

\newcommand{\lp}{\left(}
\newcommand{\rp}{\right)} 
\newcommand{\abs}[1]{\left| #1 \right|}

\definecolor{cerulean}{rgb}{0., 0.62,0.9}

\usepackage[T1]{fontenc} 

\title{Freeze-in, glaciation, and UV sensitivity from light mediators}
\author{Nicolas Fernandez,}
\author{Yonatan Kahn,}
\author{and Jessie Shelton}

\affiliation{ Illinois Center for Advanced Studies of the Universe, University of Illinois at Urbana-Champaign, Urbana, IL 61801, USA}
\affiliation{Department of Physics, University of Illinois at Urbana-Champaign, Urbana, IL 61801, USA}

\emailAdd{nicofer@illinois.edu}
\emailAdd{sheltonj@illinois.edu}
\emailAdd{yfkahn@illinois.edu}

\abstract{
Dark matter (DM) freeze-in through a light mediator is an appealing model with excellent detection prospects at current and future experiments.  Light mediator freeze-in is UV-insensitive insofar as most DM is produced at late times, and thus the DM abundance does not depend on the unknown early evolution of our universe.  However the final DM yield retains a dependence on the initial DM population, which is usually assumed to be exactly zero.  We point out that in models with light mediators, the final DM yield will also depend on the initial conditions assumed for the light mediator population.  We describe a class of scenarios we call ``glaciation'' where DM freezing in from the SM encounters a pre-existing thermal bath of mediators, and study the dependence of the final DM yield on the initial temperature of this dark radiation bath.
To compute DM scattering rates in this cosmology, we derive for the first time an exact integral expression for the Boltzmann collision term describing interactions between two species at different temperatures. 
We quantify the dependence of the DM yield on the initial dark temperature and find that it can be sizeable in regions near the traditional (zero initial abundance) freeze-in curve.  We generalize the freeze-in curve to a glaciation band, which can extend as much as an order of magnitude below the traditional freeze-in direct detection target, and point out that the DM phase space distribution as well as the yield can be strongly dependent on initial conditions.
}

\begin{document} 
\maketitle
\flushbottom

\section{Introduction}
\label{sec:intro}

The hypothesis of thermal contact between dark matter (DM) and the Standard Model (SM) is a powerful organizing principle for predictive and testable models of DM. The most common such paradigm is thermal freeze-out \cite{kolb2018early}, where DM is in thermal equilibrium with the SM for some period before the expansion rate of the universe exceeds the DM annihilation rate. In this case the relic abundance of DM is a remnant of the original thermal population, and is ``UV-insensitive'' in the sense that it is only physics at late times that sets the DM abundance. An alternate paradigm is freeze-in \cite{McDonald:2001vt,Choi:2005vq,Petraki:2007gq,Hall:2009bx,Bernal:2017kxu}, where the relic abundance of DM is gradually built up through rare processes that produce DM from the SM thermal plasma. In this scenario the DM never attains thermal equilibrium with the SM and thus predictions for its abundance necessarily retain some dependence on initial conditions; however, when the mediating interactions are renormalizable, the DM production rate peaks at late times, resulting in a weaker but still valuable form of UV-insensitivity. The freeze-in paradigm is especially appealing as a target for direct detection \cite{Alexander:2016aln,Battaglieri:2017aum}, because if the mediator of the interaction is light (for example, a kinetically-mixed dark photon), even weak DM-SM interactions are enhanced at low velocities, leading to promising sensitivity at current and future terrestrial experiments \cite{Essig:2011nj,Graham:2012su,Essig:2015cda,Lee:2015qva,Hochberg:2015pha,Hochberg:2015fth,Alexander:2016aln,Derenzo:2016fse,Hochberg:2016ntt,Battaglieri:2017aum,Essig:2017kqs,Cavoto:2017otc,Hochberg:2017wce,Essig:2018tss,Geilhufe:2018gry,Hochberg:2019cyy,Trickle:2019nya,Griffin:2019mvc,Coskuner:2019odd,Geilhufe:2019ndy,Catena:2019gfa,Blanco:2019lrf,Kurinsky:2019pgb,Berlin:2019uco,Kurinsky:2020dpb,Griffin:2020lgd,Radick:2020qip,Gelmini:2020xir,Trickle:2020oki,Du:2020ldo,Hochberg:2021pkt,Knapen:2021run,Knapen:2021bwg,Griffin:2021znd,Hochberg:2021ymx,Lasenby:2021wsc,Essig:2012yx,Tiffenberg:2017aac,Romani:2017iwi,Crisler:2018gci,Agnese:2018col,Agnes:2018oej,Settimo:2018qcm,Akerib:2018hck,Abramoff:2019dfb,Aguilar-Arevalo:2019wdi,Aprile:2019xxb,Barak:2020fql,Arnaud:2020svb,Amaral:2020ryn,Hochberg:2021yud} (see Ref.~\cite{Kahn:2021ttr} for a review of recent progress).

In models of ``traditional'' IR-dominated freeze-in where the DM does not interact appreciably after it is produced, the residual UV sensitivity amounts to a constant offset in the DM yield for a given parameter point, as we briefly discuss.  However, producing DM via a light mediator necessarily implies that DM interactions with mediator particles can give rise to DM-number-changing processes at cosmologically-interesting rates.  A cosmological population of dark {\em mediators} can therefore substantially affect the final DM number density that results from a particular coupling to the SM. Here we quantitatively assess the UV sensitivity that arises in models with different initial conditions for the light mediator, and demonstrate that different initial conditions for the dark sector can give rise to very different cosmological histories for the same couplings.
We consider the simple and generic scenario when this population is \emph{thermal}, i.e., in kinetic equilibrium at a temperature $\tilde T$, which in general will differ from the SM temperature $T$.

There is a substantial body of literature studying the interplay between freeze{-}in and freeze{-}out processes in determining the final relic abundance of DM in hidden sectors with light mediators \cite{Chu:2011be,Bernal:2015ova,Krnjaic:2017tio,Berlin:2017ftj,Berger:2018xyd,Evans:2019vxr,Hambye:2019dwd, Du:2020avz,Tapadar:2021kgw, Hryczuk:2021qtz}.  These studies consider the case where the energy density in the dark radiation bath is built up entirely from the energy injected from the SM. 
The novel point we focus on here is the qualitatively new sensitivity of the final DM relic abundance to the {\em initial} dark sector population (DM and light mediators), which we parameterize through an initial temperature ratio $\xi_i = \tilde T_{i}/T_{i}$; previous work corresponds to setting $\xi_i =0$. We call freeze-in into a {\em pre-existing} thermal bath ``glaciation''.
Taking a Dirac fermion $\chi$ interacting with a light kinetically-mixed dark photon $Z_D$ ($m_{Z_D} \ll m_\chi$) as our benchmark model for the dark sector, we establish the regions of parameter space where both traditional freeze-in and glaciation are self-consistent descriptions of the theory.
We demonstrate that for larger values of the model couplings, the energy injection from the SM overwhelms the initial conditions and predictions are UV-insensitive, while for values near the traditional freeze-in curve, the realized DM abundance can depend sensitively on the initial temperature ratio.  Our results substantially clarify the theoretical status of the freeze-in curve as a  target for direct detection experiments, and motivate an expanded ``glaciation band'' which can extend up to an order of magnitude below the freeze-in cross section. 

This paper is organized as follows. In Sec.~\ref{sec:freezein}, we review the traditional freeze-in paradigm with a light kinetically-mixed mediator and zero initial abundance, and show that the assumption of no self-interactions is valid up to a maximum value of $\alpha_D$. In Sec.~\ref{sec:glaciation}, we introduce the thermalized dark sector population and set up and solve the Boltzmann equations relevant for the more general glaciation scenario. As a consequence of our analysis, we derive for the first time an exact expression for the collision term describing interactions between two populations at different temperatures.  These Boltzmann equations assume that the injected DM achieves rapid kinetic equilibrium with the SM, and in this section we delineate the parameter space where this assumption is valid.
We present our results in Sec.~\ref{sec:results}, including the implications for direct detection experiments searching for DM-electron scattering. We conclude in Sec.~\ref{sec:conclusions}. Details of our solutions to the Boltzmann equations are given in Appendices~\ref{sec:collision} and~\ref{app:CS}.

\section{Freeze-in with a light dark photon mediator}
\label{sec:freezein}

DM lighter than 10 GeV is strongly constrained by energy injection constraints from the cosmic microwave background (CMB) \cite{Slatyer:2015jla,Planck:2018vyg}. The freeze-in mechanism is a generic way to avoid excess late-time DM annihilation, since there is never enough DM for the annihilation process to be active, and thus there is no need to appeal to a velocity-suppressed annihilation cross section which implies constraints on the spin and parity of the DM or mediator. 

\subsection{Benchmark dark photon model}
A standard benchmark model which realizes the ``traditional'' freeze-in scenario contains Dirac fermion DM $\chi$ that interacts with a dark photon, $Z_D$, with dark gauge coupling $g_D$.  The dark photon communicates with the SM through kinetic mixing
with SM hypercharge \cite{Galison:1983pa,Holdom:1985ag},
\beq
\mathcal L_{mix} = - \frac{\epsilon}{2 \cos\theta_W} \hat Z_{D\mu\nu} \hat B^{\mu\nu} \,.
\eeq
We take the dark photon to have a small but non-zero mass $m_{Z_D}$, which for simplicity we consider to arise from a St\"uckelberg mechanism \cite{Stueckelberg:1938zz,Feldman:2007wj}. We will typically be interested in $\chi$ masses below 1 GeV.

In the regime where DM never attains thermal equilibrium with the SM, the portal coupling $\epsilon$ is very small, and the couplings of $Z_D$, $Z$ to SM fermions $f$ and DM are to an excellent approximation given by:
\begin{eqnarray}\nonumber
\mathcal{L}  &\supset& g_{Z_D\chi} \bar \chi \gamma_\mu \chi Z_D^\mu  + g_{Z\chi} \bar \chi \gamma_\mu \chi Z^\mu +\sum_f g_{Z_D f} \bar f \gamma_\mu f Z_D^\mu ,\\
g_{Z_D f} & \approx& -\epsilon\dfrac{g}{\cos\theta_W}\left(\tan\theta_W\dfrac{m_Z^2}{m_Z^2-m_{Z_D}^2}(T_3\cos^2\theta_W-Y\sin^2\theta_W)+Y\tan\theta_W\right),\\\nonumber
g_{Z_D\chi}&\approx& g_D,\,\\
\label{eq:zcouplfinal}
g_{Z\chi}&\approx& \epsilon g_D\tan\theta_W\dfrac{m_Z^2}{m_Z^2-m_{Z_D}^2},
\end{eqnarray}
while the coupling of the $Z$ boson to SM fermions is to leading order unaltered.
This DM model can thus be described at the Lagrangian level by four parameters,
which we will take to be $\alpha_D,\epsilon,m_\chi$ and $m_{Z_D}$ where $\alpha_D = g_D^2/(4\pi)$. However, when $m_{Z_D} \ll m_\chi$, the regime of greatest interest for direct detection, the dark cosmological history as well as the resulting direct detection signals are largely insensitive to the specific value of the dark photon mass. In the limit $m_{Z_{D}} \ll m_{Z}$ of interest, the couplings of the $Z_D$ reduce to the simpler expressions  $g_{Z_D f} \approx -\epsilon  e Q_{f}$, $g_{Z_D\chi} \approx g_{D}$ and $g_{Z \chi} \approx \epsilon g_D\tan\theta_W$. 

\subsection{Traditional freeze-in: review}
With the mass hierarchy $m_{Z_D} \ll m_\chi$, and the absence of any additional dark sector species, freeze-in is UV-insensitive in the following sense. DM is produced from annihilation of SM particles in the thermal plasma, ${\rm SM} + {\rm SM} \to \chi + \bar{\chi}$.
The DM abundance grows monotonically with time, reaching a maximum once the temperature drops below either $m_\chi$ or $m_e$, whichever is larger: in the former case, DM production becomes Boltzmann-suppressed at $T = m_\chi$, and in the latter case, the abundance of SM particles coupling to the dark photon becomes Boltzmann-suppressed after positron annihilation and plasmon decays become more important \cite{Dvorkin:2019zdi, Chang:2019xva, Dvorkin:2020xga} for the production of DM. The lightness of the dark photon is crucial here, allowing the $s$-channel annihilation to be dominated by the lightest mass scale (or lowest temperature) in the problem, rather than (say) by the mass of a new heavy mediator. Since DM production originates from  the thermal SM plasma and most of the DM is produced at late times, this mechanism is insensitive to the unknown early history of our universe.
The parameters required to achieve the observed relic abundance are \cite{Battaglieri:2017aum}
\begin{equation}
\epsilon^2 \alpha_D \simeq \begin{cases}10^{-24}, \qquad  \quad \ m_\chi > m_e \\ 
10^{-24} \frac{m_e}{m_\chi}, \qquad   m_{\chi} < m_e.\end{cases}
\label{eq:FI_param}
\end{equation}
Including DM production through the plasmon channel decreases the couplings required to achieve the freeze-in relic density by up to an order of magnitude for $m_\chi < m_e$ \cite{Dvorkin:2019zdi, Dvorkin:2020xga}, but the above estimates are sufficient since we primarily focus on the regime $m_\chi > m_e$ in this paper. As mentioned in the Introduction, the hidden UV sensitivity in this nominally UV-insensitive scenario is the choice of initial DM abundance, which is customarily taken to be zero. We refer to this scenario as ``traditional'' freeze-in.
We now show that the only effect of a nonzero initial $\chi$ abundance is a simple offset in the late-time relic abundance, rendering this residual UV-sensitivity rather trivial. Assuming for simplicity that $e^+ e^- \to \chi \bar{\chi} $ is the only process which populates the dark sector aside from any primordial abundance, the Boltzmann equation relating the DM abundance $n_\chi$ to the electron abundance $n_e$ is
\begin{equation}
\dot{n}_\chi + 3 H n_\chi = 2 \langle \sigma v \rangle n_e^2,
\label{eq:FI_simple}
\end{equation}
where $H \equiv \dot{a}/a$ is the Hubble parameter and $\langle \sigma v \rangle$ is the thermally-averaged annihilation cross section. Changing variables to the comoving yield $Y_\chi = n_\chi/s$ (where $s$ is the entropy density) and to the dimensionless time variable $x = m_\chi/T$, we have
\begin{equation}
H x \frac{dY_\chi}{dx} = \frac{2\langle \sigma v \rangle n_e^2}{s}.
\label{eq:nx_simple}
\end{equation}
Consider first the regime where $x \ll 1$, and assume that $m_\chi > m_e$. In that case, for annihilation through a light $Z_D$, $\langle \sigma v \rangle \sim \pi \alpha_D \epsilon^2 \alpha / T^2 \propto x^2$ since there are no Boltzmann suppressions or kinematic endpoints. Electrons are always relativistic, so $n_e \propto T^3 \propto 1/x^3$, and similarly, $s \propto T^3 \propto 1/x^3$. By assumption, freeze-in is taking place during radiation domination, so $H \propto 1/x^2$. Collecting the $x$ dependence, we find
\begin{equation}
\frac{dY_\chi}{dx} = {\rm const.}
\end{equation}
which has the trivial solution $Y_\chi = Y_0 +  {\rm const.} \times x$. So the effect of a primordial abundance $Y_0$ is simply to offset the linear growth of $Y_\chi$, which will push the slope of $Y_\chi(x)$ to smaller values (in other words, smaller couplings $\epsilon^2\alpha_D$) to achieve the same DM abundance when freeze-in turns off around $x \sim 1$.

\subsection{Self-consistency of traditional freeze-in}
The above analysis has made an implicit assumption that dark particles, once produced from the SM, subsequently free-stream without further interaction. In some portions of our four-dimensional parameter space, this assumption does indeed hold. In other parts of parameter space, however, interactions of the injected dark matter particles, both with each other and with the light dark mediator particle, are important and can lead to sizable impacts on the DM phase-space distribution or even the final relic abundance. 
We estimate here the regime of validity of this traditional freeze-in treatment by requiring that a frozen-in particle does not undergo further scattering after production.

Writing the elastic scattering rate between frozen-in DM particles as $n_{\chi} \langle \sigma v \rangle_{\rm{el}}$, we have from Eq.~(\ref{eq:nx_simple}) that $n_{\chi}\approx n^{2}_{f} \langle \sigma v \rangle/ H$ where $f$ is an SM fermion. Therefore, by simply imposing that the elastic scattering rate be smaller than the Hubble rate, we have the condition
\beq
n_{f}\sqrt{\langle \sigma v \rangle_{\rm{el}} \langle \sigma v \rangle} \lesssim H.
\eeq
Interestingly, the rate is given by the number density of the particles annihilating into DM and by an \emph{effective} cross section which is the geometric mean of the scattering and the annihilation cross sections. We estimate the thermally-averaged scattering cross section as $\langle \sigma v \rangle_{\rm{el}} \approx \pi \alpha^{2}_{D}/T^{2}$, and thus $n_{f}\sqrt{\langle \sigma v \rangle}\approx \sqrt{2 \alpha \alpha_{D} \epsilon^{2}/(3 \pi^{3})} T^{2}$ for $T \gtrsim m_{\chi}$. The bound for the combination of couplings $\alpha_{D}$ and $\epsilon$ for negligible elastic scattering would be 
\beq
\label{eq:alphaepsilonmaxNoScattering}
\alpha^{3}_{D} \epsilon^{2} \lesssim 2.2\times 10^{3} \left( \dfrac{g_{*\rho}}{10} \right) \left( \dfrac{m_{\chi}}{\MPl} \right)^{2} \,,
\eeq
where $g_{*\rho}$ is the effective number of relativistic degrees of freedom related to the energy density. To get the correct freeze{-}in abundance (Eq.~(\ref{eq:FI_param})), we need $\epsilon^{2}\alpha_{D} \approx 3.5 \times 10^{-24}$. Therefore, in this case our estimation for the maximum value of $\alpha_{D}$ self{-}consistent with the traditional freeze{-}in mechanism is
\beq
\alpha^{\rm{max}}_{D}\approx 10^{-8} \left( \frac{m_{\chi}}{\MeV} \right) \qquad \text{(self-consistent traditional freeze-in)}.
\eeq 
For other values of $\epsilon$ and $\alpha_D$, both self-scattering and self-annihilations are important, and the initial condition dependence becomes more involved.  We turn to this region of parameter space in the following section.

\section{Freeze-in into a pre-existing thermal bath}
\label{sec:glaciation}

A more interesting type of UV sensitivity, with rich accompanying dynamics, arises when there is a pre-existing population of a dark sector containing $\chi$, rather than simply a non-interacting primordial DM abundance.  

As DM is injected into this dark thermal bath, it will exchange kinetic energy with bath particles. Further, annihilations within the dark sector may begin to deplete the DM abundance, in sharp contrast to the monotonic increase in the traditional freeze-in scenario described above.  We parameterize the initial conditions on this dark thermal bath through an initial temperature ratio $\xi_i \equiv \tilde T_i/T_i$.
Our regime of interest is $\xi_i<1$, and therefore the Hubble rate is always dominated by the SM energy density, $H(T,\tilde T)\approx H(T)$.

\subsection{Boltzmann equations}
\label{sec:boltz}
For a kinetically mixed $Z_D$, the dominant source of energy injection into the hidden sector is through DM pair production. Since this injection can easily occur after DM has already departed from full chemical equilibrium, it is important to track how much of this energy is converted into the shared dark sector temperature $\tilde T$ and how much remains sequestered as rest mass. In other words, the energy density of the hidden sector, $\rho_{\rm HS} = \rho_{Z_{D}} + 2\rho_{\chi}$, as well as the number density of DM, $n_{\rm DM}= 2n_{\chi}$, are determined by the DM chemical potential  $\mu$ as well as the hidden sector temperature $\tilde T$.\footnote{In what follows, we assume there is no dark matter asymmetry, i.e $n_\chi = n_{\bar{\chi}}$.}

The corresponding Boltzmann equations can be written as
\barray
& &\dot\rho_{\rm HS} + 3 H \left( \rho_{\rm HS} + P_{\rm HS}\right) = \sum_{f} \langle \sigma v E\rangle_{\rm{fi }\, } n_{f}^2(T) + \langle \Gamma E \rangle_{Z} n_{Z}(T) \label{eq:BE_energy} \\ 
& &\dot n_{\rm DM} + 3 H n_{\rm DM} = -\frac{1}{2}\langle \sigma v\rangle_{\rm{fo}} (n_{\rm DM}^2 - n_{\rm eq}^2(\tilde T)) + 2\sum_{f} \langle \sigma v \rangle_{\rm fi} n_{ f}^{2}(T) + 2 \langle \Gamma \rangle_{Z} n_{Z}(T) \label{eq:BE_number}\, ,
\earray
where the sums run over SM fermions $f$ and $P_{\rm HS}= P_{Z_{D}} + 2 P_{\chi}$ is the pressure of the hidden sector.  The collision terms appearing in Eq.~(\ref{eq:BE_energy}), $\mathcal{C}_{f\bar{f} \rightarrow \chi \bar{\chi}}^{\rho}(T) = n_{f}^2(T) \langle \sigma v E \rangle$ and $\mathcal{C}_{Z\rightarrow \chi \bar{\chi}}^{\rho}(T) = \langle \Gamma E \rangle_{Z} n_{Z}(T)$, govern the injection of energy into the HS from DM pair production, where $E=E_{1}+E_{2}$ ($E=E_Z$) is the total energy of the annihilating fermions (decaying $Z$ boson). In the regime of primary interest to us, the first term, describing production from SM fermion annihilations, dominates over the second term, which indicates the contribution from $Z$ decays. 

The collision terms appearing in Eq.~(\ref{eq:BE_number}) include the effect of DM annihilations within the hidden sector as well as the injection of DM from the SM.  The specific expressions for the various thermally-averaged quantities appearing in the collision terms are given in Appendix~\ref{app:collision_terms}.  The analogous Boltzmann equation for the SM temperature (including the effect of reverse annihilations), along with the Friedmann equation giving the dependence of $H$ on $T$ and $\tilde T$, provide a closed system of equations. We solve this set of equations numerically using the dimensionless time variable $x=m_\chi/T$ with initial condition $x_i = 10^{-2} \xi_{i}$. This initial condition defines the initial temperature ratio $\xi_i$ at the SM temperature $T_i=10^{2} m_\chi/\xi_i$, which ensures  $\tilde T_{i} = 10^{2} m_{\chi}$ for all values of $\xi_i$, and thus makes sure we set initial conditions early enough to capture the correct DM evolution for all cases.  

These Boltzmann equations have made one major assumption: that the DM number density (and thus energy density and pressure) can be described entirely in terms of $\mu$ and $\tilde T$, or in other words, that DM can always be taken to be in kinetic equilibrium with the mediator bath.  This is the opposite limit from traditional freeze-in, where after production the DM phase-space distribution evolves only through redshifting.  The description in terms of $\mu$ and $\tilde T$ is valid when DM produced via freeze-in rapidly reaches kinetic equilibrium with the dark radiation bath, which holds over the parameter space of primary interest to us; we demonstrate the self-consistency of this assumption in Sec.~\ref{sec:kineq}.  

We can gain some intuition about this system of equations by first considering the situations where the dark sector is in internal chemical equilibrium, in which case Eq.~(\ref{eq:BE_energy}) for $\rho_{HS}$ is the only necessary equation to solve.
In this case the dark sector temperature $\tilde T$  evolves non-adiabatically with scale factor once the  rate of energy injection from the SM becomes comparable to the rate of energy dilution owing to the expansion of the universe \cite{Cheung:2010gj, Chu:2011be, Krnjaic:2017tio, Berger:2018xyd, Evans:2019vxr}.
Once the energy injection from the SM shuts off, the energy density in both sectors resumes adiabatic evolution.  Thus during the time that the dark sector is in chemical equilibrium, the temperature evolution during the non-adiabatic period, which we will refer to as the ``leak-in'' phase for clarity, follows a cosmological attractor solution $\tilde T_{LI}(a)$ \cite{Evans:2019vxr}: given $\tilde T \ll T$ and a collision term $C_E (T)\propto \epsilon^2 \alpha_D$ describing the rate of energy transfer into the HS, $\tilde T_{LI}(a)$ is entirely fixed in terms of the SM temperature, with $\tilde T_{LI}(a) \propto (\epsilon^2 \alpha_D)^{1/4}$.  When $C_E\propto T^5$, as is generic in the absence of mass thresholds, the resulting leak-in solution gives $\tilde T_{LI}(a)\propto a^{-3/4}$. Hidden sectors with $\tilde T(a_i) > \tilde T_{LI}(a_i)$ evolve adiabatically until $\tilde T (a)=\tilde T_{LI} (a)$ and subsequently follow the leak-in solution, while hidden sectors with $\tilde T(a_i) < \tilde T_{LI}(a_i)$ see their temperature rapidly rise up to the attractor solution. The approximate scaling of the attractor solution, normalized to the SM temperature and written in terms of temperature instead of scale factor for future convenience, is
\begin{equation}
\label{eq:xiLI}
    \xi_{LI}(T) \approx 10^{-2} \left( \alpha \alpha_{D} \epsilon^2  \MPl / T \right)^{1/4},
\end{equation}
where $\alpha = e^2/(4\pi)$ is the QED coupling.

The existence of this IR-dominated attractor solution helps mitigate the sensitivity of the DM relic abundance to the initial value of $\xi_i$, since sectors with $\xi_i < \xi_{LI} (T_i)$ will trend toward to the attractor temperature ratio $\xi_{LI}(T)$.  However as chemical equilibrium is lost within the dark sector, it is necessary to keep more careful track of how much the energy injected from the SM  is distributed. 
The system will leave the attractor solution once any of the following conditions are met: (i) the energy injection from the SM shuts off; (ii) the HS departs from chemical equilibrium; (iii) the energy density in the HS is dominated by matter, rather than radiation.  To understand in detail which of these conditions is most relevant for any given parameter point, we need to numerically solve the full system described by Eqs.~(\ref{eq:BE_energy}) and~(\ref{eq:BE_number}), to which we turn in the next section.

Finally, sufficiently large portal couplings will thermalize the dark sector with the SM. In other words, at a sufficiently large value of $\epsilon$, the dark sector reaches $\tilde{T}=T$ for a given $\alpha_{D}$. The attractor solution gives a quick way to estimate when thermalization occurs.  On the attractor, the dark temperature is given by $\tilde{T}^{4}=C \MPl \epsilon^{2} \alpha_{D}  T^{3}$, where $C$ is a dimensionless constant. Thus setting $\tilde{T}=T= C \MPl \epsilon^{2} \alpha_{D}$ lets us estimate when the two sectors thermalize. We are interested in the temperature range $T> m_{\chi}$, and therefore the value of $\epsilon$ at which the two sectors thermalize is
\begin{align}
\label{eq:epsilon_thermal}
\epsilon^{\rm thermal} \approx 1.2 \times 10^{-5} \left(\frac{10^{-7}}{\alpha_{D}}\right)^{1/2} \left( \frac{m_{\chi}}{\MeV}\right)^{1/2} \,.
\end{align}

\subsection{Kinetic equilibration}
\label{sec:kineq}

The Boltzmann equations given in Eqs.~(\ref{eq:BE_energy})-(\ref{eq:BE_number}) are a good description of the system as long as the DM produced from out-of-equilibrium interactions with the SM rapidly reach kinetic equilibrium with the dark radiation bath.
A $\chi \bar{\chi}$ pair injected into a dark thermal bath of temperature $\tilde{T}$ can 
interact with both the DM and dark photons within the bath.  Kinetic equilibrium can be obtained through scattering of injected DM with bath DM particles via a $t$-channel $Z_D$, as well as the Compton scattering of injected DM from a $Z_D$ in the bath.
The injected DM can also  approach chemical equilibrium through annihilating with bath particles, via the $t$-channel process $ \chi \bar{\chi} \to Z_D Z_D$. In the regime of interest $T\gg \tilde{T}$, the Hubble rate is determined by the SM temperature, meaning that $H \propto T^2/M_{\rm pl}$. To attain kinetic equilibrium, the momentum loss rate  of the injected DM due to scattering with some particle in the pre-existing dark thermal bath ($\chi$, $\bar{\chi}$ or $Z_{D}$) needs to be greater that the Hubble rate, i.e.,
\beq
 \Gamma_{p \, \mathrm{loss}} \equiv \langle \frac{d\Delta p^{2}}{dt} (T,\tilde T) \rangle  \frac{1}{ \langle p^{2} (T)\rangle} = \frac{n_{2 \rm eq}(\tilde{T}) \langle \sigma v  \Delta p^{2} \rangle }{\langle p^{2} \rangle} \gtrsim H(T)\,,
\eeq
where we have defined the fractional momentum loss rate with respect to the momentum $p(T)$ of an injected DM particle in a Lorentz-invariant way. To compute this rate we derive new exacts result for collision terms describing the scattering of particles at two different temperatures, given in~Apps.~\ref{sec:num_den}--\ref{app:ptransfer}. 
\begin{figure}[!t]
\center
\includegraphics[width=0.48\textwidth]{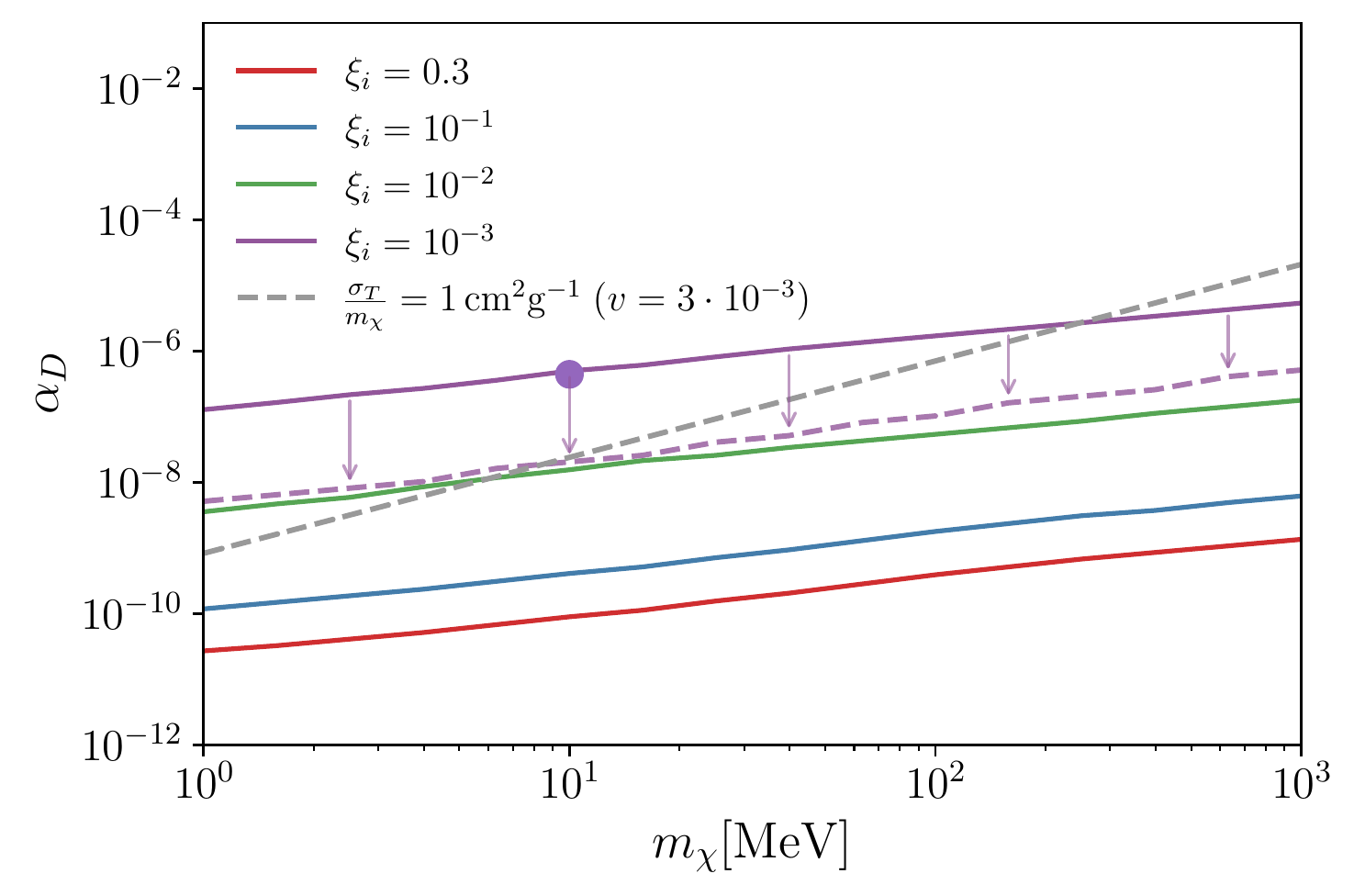}
\includegraphics[width=0.48\textwidth]{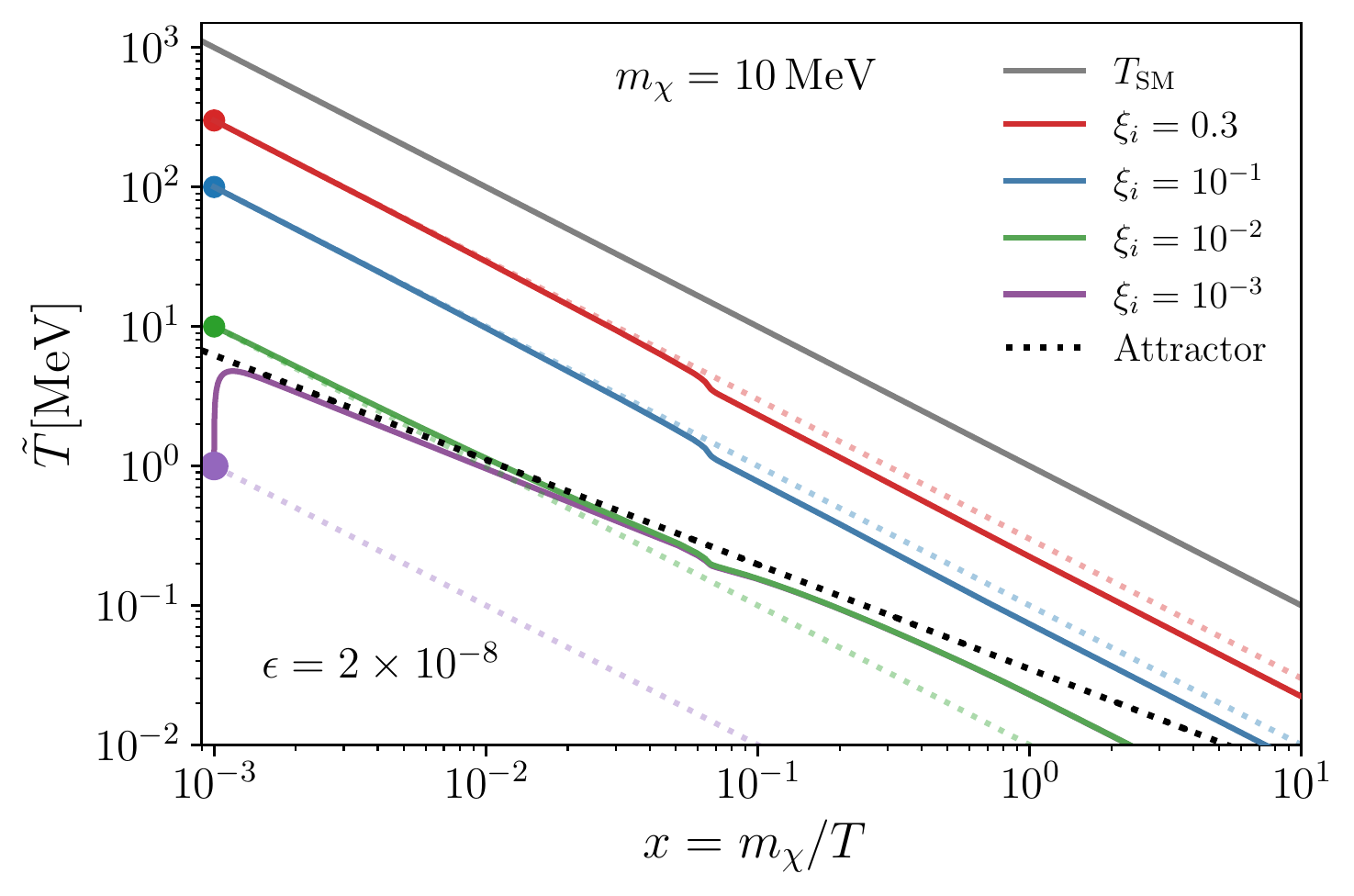}
\caption{\textbf{Left panel:} Minimum value of $\alpha_D$ required for the rapid kinetic equilibration of DM, as a function of $m_\chi$.  Solid colored lines show results for a range of fixed initial temperature ratios  $\xi_i$, under the (conservative) assumption that the HS temperature evolves adiabatically.  For $\xi_{i} = 10^{-3}$ (dashed purple line) we show the case when there is energy injection, i.e.\ non-adiabatic evolution, which weakens the constraint on $\alpha_D$.  The dashed gray line shows the maximum value of $\alpha_D$ allowed by requiring the DM transfer cross-section $\sigma_T$ for self-interactions \cite{Colquhoun:2020adl} to satisfy $\sigma_T/m_\chi <1\,\mathrm{cm}^2/$g at cluster-scale velocities $v = 3\times 10^{-3}$ \cite{Tulin:2017ara}. \textbf{Right panel:} Solid lines show the evolution of the hidden temperature for $\alpha_{D}=1.1 \times 10^{-7}$ and $\epsilon=2\times 10^{-8}$ (yellow dot in Fig.~\ref{fig:evol}) for different initial temperature ratios. Faded dotted lines shows the adiabatic evolution of the HS temperature.}
\label{fig:K_eq}
\end{figure}

First consider the case when $\tilde T$ evolves adiabatically, and therefore $\xi$ is constant (up to mass thresholds). Fig.~\ref{fig:K_eq} shows the minimum values of $\alpha_D$ for which the assumption of rapid kinetic equilibrium is satisfied for a given fixed $\xi$ (solid lines). Notice as the hidden temperature gets closer to the SM temperature, smaller values for $\alpha_{D}$ are needed to obtain rapid kinetic equilibrium in the dark sector. This can be understood from the fact that as the hidden temperature increases, the number density of bath particles increases as well, giving higher interaction rates. On the other hand, when the hidden temperature is significantly less that the SM temperature, there will be fewer interactions and a bigger interaction coupling is needed for the injected $\chi$ to efficiently lose its momentum. Finally, the gray dashed line shows conservative constraints on $\alpha_D$ coming from measurements of halo ellipticities \cite{Feng:2009mn,Agrawal:2016quu} or relaxation of the halo profiles of galaxy groups and clusters \cite{Sagunski:2020spe} (see  Ref.~\cite{Tulin:2017ara} for a review of self-interacting DM constraints). 
This fixed-$\xi$ estimate can be overly conservative, however, depending on the value of $\epsilon$, as it neglects the effect of energy injection from the SM on the dark temperature. In the right panel of Fig.~\ref{fig:K_eq} we show the evolution of $\tilde T$ for a range of initial $\xi_i$ and compare to the attractor solution corresponding to a particular $\alpha_D,\epsilon$ pair (also used in Fig.~\ref{fig:evol} below).  The larger two initial temperature ratios (red and blue lines) begin above the attractor solution (dotted black) and redshift adiabatically down, while the initially underabundant purple curve rises up rapidly to the attractor.  Meanwhile the green curve redshifts down until it meets the attractor, after which it follows the attractor solution.  The effect of the QCD phase transition is visible at $x\sim 0.07$,  where the approximate attractor solution does not account for this effect. Neglecting the SM energy injection is thus an excellent approximation for the red and blue lines but  underestimates the HS temperature and therefore the scattering rate for the green and especially the purple lines, for which $\xi < \xi_{LI}(a; \epsilon, \alpha_D)$ for some $a$.  The impact of this non-adiabatic evolution on the requirement of kinetic equilibration is illustrated with the dashed purple line in Fig. ~\ref{fig:K_eq}, which shows the minimum values of $\alpha_D$ that give rapid kinetic equilibration, given the attractor solution corresponding to $\epsilon= 2\times 10^{-8}$. 

Further details about the calculation of kinetic equilibration are given in Appendix~\ref{app:IKE}.  For the values of $(m_\chi,\epsilon, \alpha_D)$ of interest in this work, rapid kinetic equilibration is a good approximation in a substantial portion of parameter space, and in particular the portion of parameter space that displays interesting dependence on initial conditions.

\section{Results}
\label{sec:results}

\begin{figure}[!t]
\centering
    \includegraphics[width=0.495\textwidth]{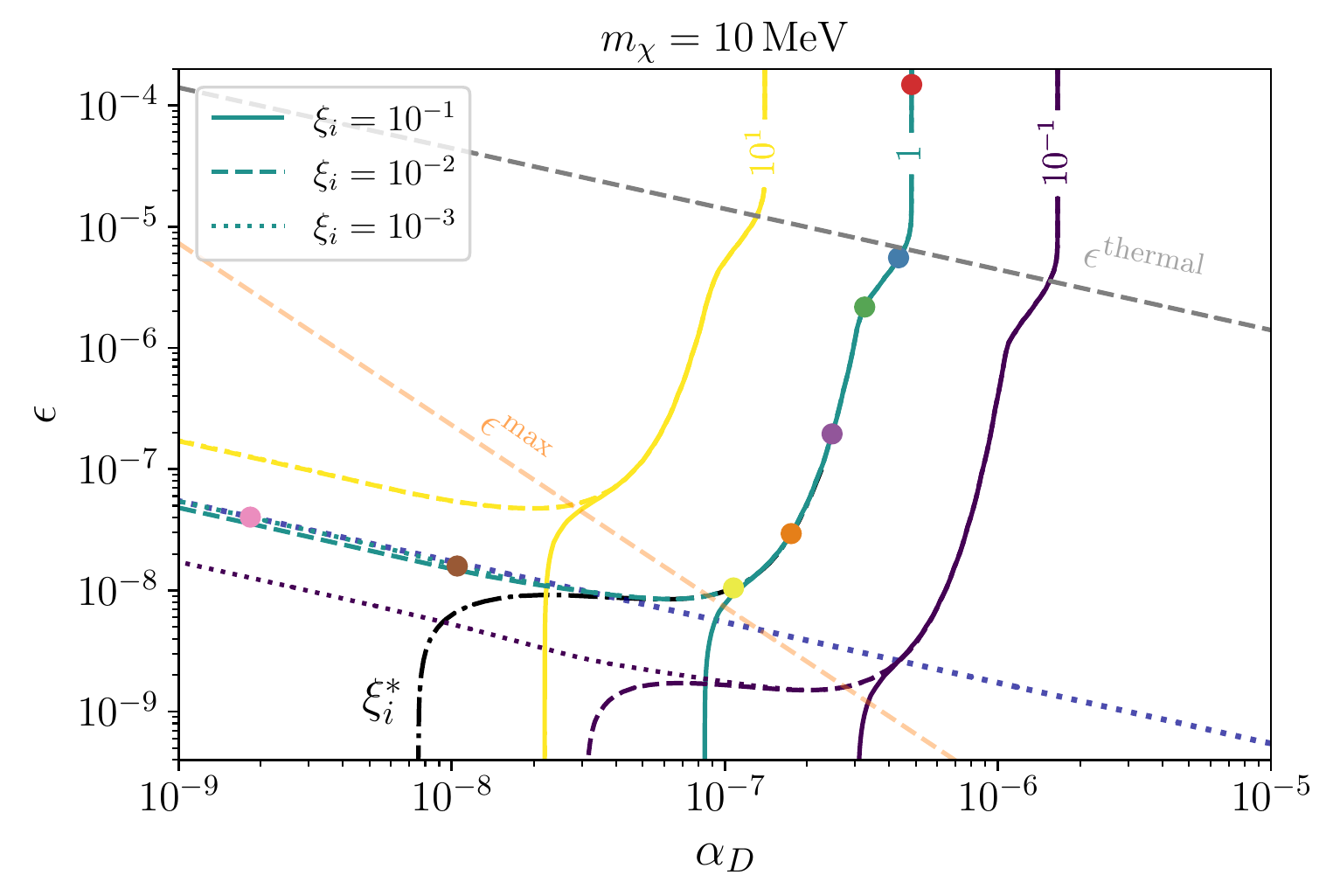}
    \includegraphics[width=0.495\textwidth]{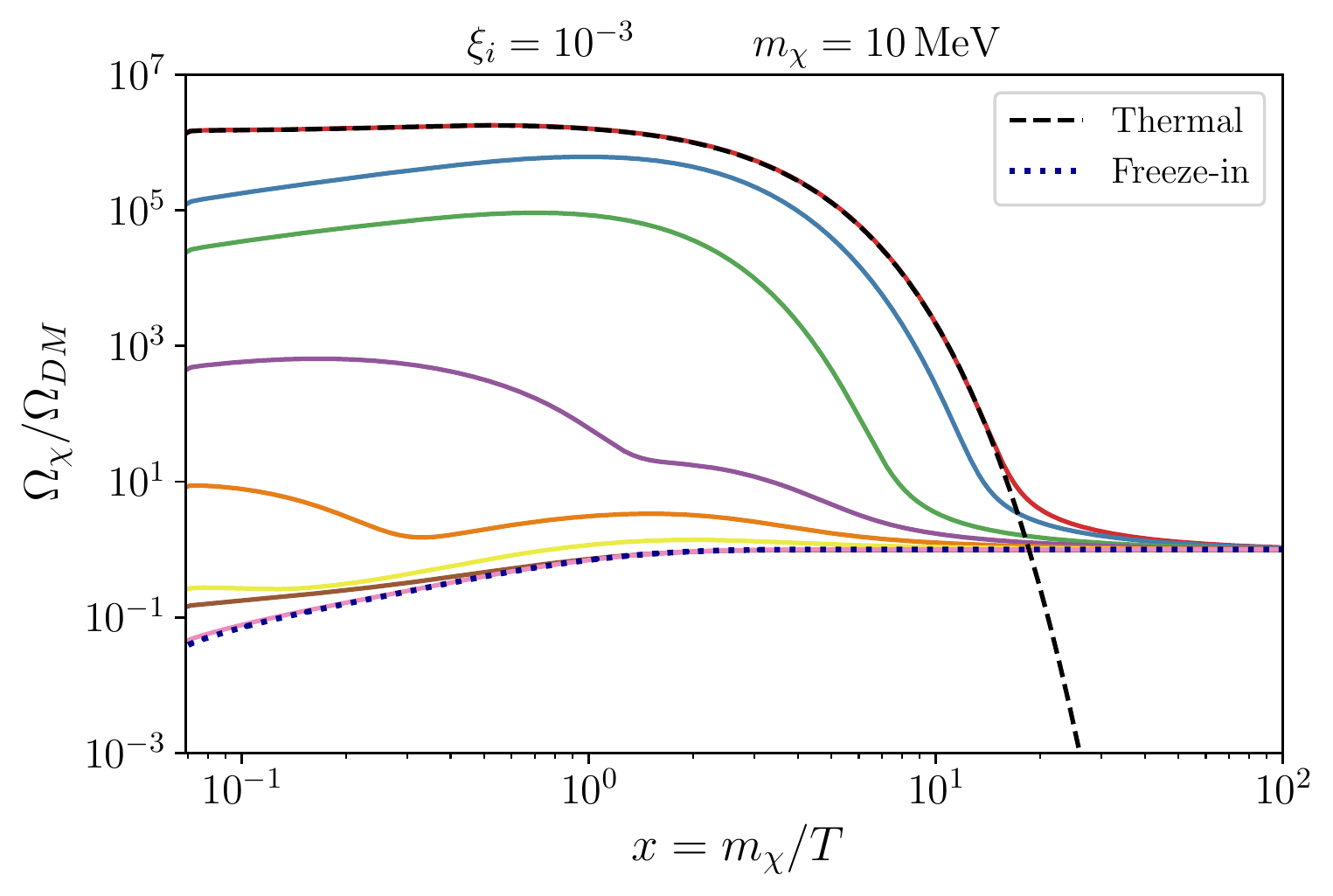}
\caption{\textbf{Left panel:} Contours of $\Omega_{\chi}/\Omega_{\rm DM}$ in the $\alpha_{D}$-$\epsilon$ plane for different initial temperature ratios. The gray dashed line corresponds to the maximum value of $\epsilon$ for a given $\alpha_{D}$, such that the hidden sector reaches thermal equilibrium with the SM (\ref{eq:epsilon_thermal}). The dark blue dotted line indicates the traditional freeze-in production (\ref{eq:FI_param}), and the orange dashed line shows the  upper bound $\epsilon^{\rm{max}}$ as a function of $\alpha_D$ (\ref{eq:alphaepsilonmaxNoScattering}) for which elastic scattering is negligible and a frozen-in DM phase space distribution evolves only through redshifting. \textbf{Right panel:}  Evolution of the ratio $\Omega_{\chi}/\Omega_{\rm DM}$ with $x$ for each colored point shown in the left panel, where the final yield of $\chi$ matches the observed DM relic abundance.} 
\label{fig:evol}
\end{figure}

The DM number density is obtained after solving the system of equations (\ref{eq:BE_energy}) and (\ref{eq:BE_number}). To develop some intuition for the strength of the couplings needed to obtain the correct DM relic abundance, we first explore the parameter space as a function of the initial temperature ratio $\xi_i$. We show the results in Fig.~\ref{fig:evol} for $m_{\chi} = 10 \ \MeV$ and different initial temperature ratios. The left panel in Fig.~\ref{fig:evol} shows contours of $\Omega_{\chi}$ (normalized to the observed DM relic density) in the $\alpha_D$-$\epsilon$ plane. This plot illustrates two distinct regimes at small coupling (bottom left corner):
\begin{enumerate}[I.]
    \item For small $\xi_i$ (short-dashed curves), at small couplings there is not enough DM in the hidden sector to achieve the required relic abundance through hidden-sector freeze-out alone, and instead the relic abundance is obtained through freeze-in, which implies a minimum $\epsilon$ for a given $\alpha_D$.
    \item For large $\xi_i$ (solid curves), obtaining the observed relic abundance is possible for arbitrarily small values of $\epsilon$, since the DM can freeze out entirely within the hidden sector, decoupled from the SM.
\end{enumerate}
We have checked that the approximation of rapid kinetic equilibrium, the conditions for which can be seen from Fig.~\ref{fig:K_eq}, holds for all of the parameter points shown in colored points (curves) in the left (right) panel of this figure, except the pink point (curve); the brown point (curve) lies at the boundary of the rapidly equilibrated region of parameter space along the freeze-in line.
For sufficiently large couplings, contours for different values of $\xi_i$ converge on the attractor solution described in Sec.~\ref{sec:glaciation}.  At these larger couplings, there are two qualitatively different scenarios for achieving the correct relic abundance, regardless of the initial temperature ratio. For $\epsilon$ above the gray dashed line, the hidden sector thermalizes with the SM and freeze-out obtains in the traditional way. For $\epsilon$ between the dotted blue and dashed gray lines, DM can obtain the correct relic abundance through leak-in (i.e., one phase of freeze-out during a period of non-adiabatic temperature evolution) and/or reannihilation (i.e., two distinct phases of freeze-out).
The right panel of Fig.~\ref{fig:evol} shows the evolution of the DM yield for the colored points marked in the left panel, showing the transition from freeze-out to leak-in/reannihilation\footnote{The reannihilation process \cite{Chu:2011be} occurs for the orange and purple points} to freeze-in for $\xi_{i} = 10^{-3}$.

The existence of Regime II demonstrates the UV sensitivity of freeze-in with a light mediator: these secluded freeze-out solutions are available only for some initial values of $\xi_i$, and the specific value of $\alpha_D$ that yields the correct relic abundance through secluded freeze-out depends on the specific value of $\xi_i$.  At sufficiently small $\xi_i$, however, secluded freeze-out does not occur, and the relic abundance is instead dominated by freeze-in processes.
We can understand the division between Regimes I and II straightforwardly by looking at the {\em initial} DM abundance as a function of $\xi_i$. First, let us define the comoving DM number density as $Y_{\chi} \equiv n_{\chi}/s$ as in Sec.~\ref{sec:freezein}. Then, we can express the observed DM density in a convenient way through the DM yield as
\begin{align}
\label{eq:DM_relic}
Y_{\rm DM} = 4.35 \times 10^{-7} \left( \frac{\MeV}{m_{\chi}} \right) \,,
\end{align}
which is equivalent to the standard, and more familiar, form $\Omega_{\rm DM} h^{2} = 0.12$. 
In the case of interest where the DM chemical potential is zero and its temperature $ \tilde T$ is different from the SM temperature $T$, we have
 \begin{align}
Y_{\chi} = \frac{n_{\chi}( \tilde T)}{s(T)} = \left( \frac{45 \, g_{\chi}}{4 \pi^{4}  g_{*s}} \right) \xi^{3} \tilde x^{2} K_{2}(\tilde x )  \,,
\end{align}
where $\xi=  \tilde T/T $, $\tilde x = m_{\chi}/ \tilde T $, $g_{*s}$ counts the effective relativistic degrees of freedom contributing to the entropy density, $g_\chi = 4$ for a Dirac fermion, and $K_2$ is a modified Bessel function. Therefore, the initial DM yield ($\tilde x \ll 1$) can be expressed as
\begin{align}
\label{eq:Initial_relic}
Y_{i \,\chi} = \left( \frac{45 \, g_{\chi}}{2 \pi^{4}  g_{*s}} \right) \xi_{i}^{3} \,.
\end{align}
Notice that the initial yield is independent of the DM mass, which is just the statement that DM is relativistic for $\tilde T > m_{\chi}$. As a result, the initial yield is entirely fixed by  the initial hidden-to-SM temperature ratio, $\xi_i$. On the other hand, the late-time DM yield, Eq.~(\ref{eq:DM_relic}), only depends on the DM mass. This leads to two possibilities:
\begin{itemize}
\item If $Y_{i \,\chi} > Y_{\rm DM}$, there is too much DM initially and DM needs to annihilate to reproduce the correct relic density, which is accomplished by freeze-out.\footnote{An important caveat to this argument is when there is entropy injection into the SM, e.g.\ decay of long-lived moduli, leading to a depletion of the DM relic abundance \cite{Evans:2019jcs}.}
\item If $Y_{i \,\chi} < Y_{\rm DM}$, there is too little DM initially and the DM abundance needs to build up over time, which is accomplished by freeze-in.
\end{itemize}
\begin{figure*}[!t]
\centering
\includegraphics[width=0.65\textwidth]{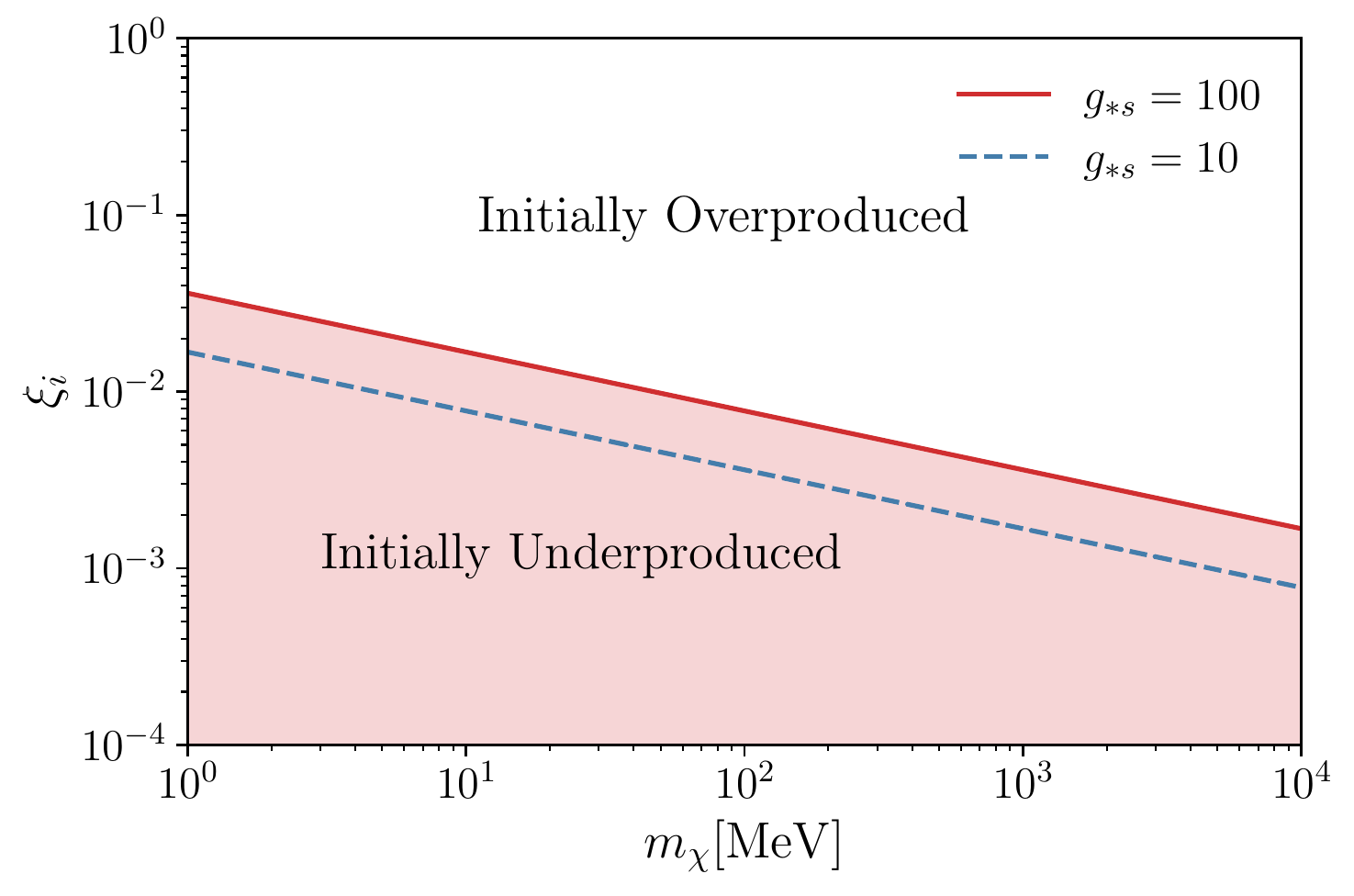}
\caption{Initial DM density as a function of the initial hidden-to-SM temperature ratio $\xi_{i}$ and DM mass $m_{\chi}$. The boundary between initial overproduction and initial underproduction (which depends on $g_{*s}$) defines the parameter space for which the freeze-out or freeze-in mechanisms are viable.}
\label{fig:xi_vs_mx}
\end{figure*}

We show in Fig.~\ref{fig:xi_vs_mx} the values of $\xi_{i}$ and $m_{\chi}$ for which the freeze-out or freeze-in mechanism is needed. We use $g_{\chi}=4$ and we show two representative values of $g_{*s}$: 100, when $T_{i} \gtrsim 200 \, \mathrm{MeV}$ (i.e.\ above the QCD phase transition) and 10 when $ T_{i} \lesssim 20 \, \mathrm{MeV}$ (below the QCD phase transition). Finally, we define the critical temperature ratio $\xi_{i}^{*}$, such that $Y_{i \,\chi}=Y_{\rm DM}$, meaning the initial yield precisely coincides with the observed relic abundance. 

These two possibilities (underproduced vs.~overproduced) map onto Regimes I and II discussed above.  However, due to the attractor solution, the ``true'' initial temperature ratio at early times will not be $\xi_i$ but rather $\xi_{LI} (T_i)$, so long as $\xi_i < \xi_{LI} (T_i)$, where the temperature evolution of the attractor solution is given in Eq.~(\ref{eq:xiLI}). Thus we need to check whether the boundary between the initially under- vs.~overproduced regimes is robust against the attractor solution for values of $\alpha_D$ and $\epsilon$ along the freeze-in curve. An initial temperature ratio $\xi_i < \xi_i^*$ will remain in the underproduced region so long as $ \xi_{LI}(T_i) < \xi_i^*$ where $T_i$ is the temperature at which $\xi_i$ is defined. The critical value $\xi^*$ at the overproduced/underproduced boundary from 
Fig.~\ref{fig:xi_vs_mx} is $\xi^* \sim 10^{-2}$ with a weak dependence on the DM mass. Along the freeze-in trajectories in Fig.~\ref{fig:evol}, $\alpha_D \epsilon^2 \simeq 10^{-24}$, so
\begin{equation}
    \xi_{LI}(T_i) < \xi_i^* \implies T_i \gtrsim 70 \ {\rm eV},
\end{equation}
a condition which is clearly required in order to have freeze-in of DM with mass greater than 1 MeV. Therefore, along the freeze-in curve seen in the left panel of Fig.~\ref{fig:evol}, the product of couplings $\alpha_D\epsilon^2$ is too small for the corresponding attractor solutions to raise these parameter points from the underproduced region to the overproduced region.  

In other words, along the freeze-in curve in $\alpha_D$-$\epsilon$ space, the initial DM production regime found in Fig.~\ref{fig:xi_vs_mx} is robust against the non-adiabatic evolution of $\tilde T$, which gives us a simple way to understand the small-$\epsilon$ behavior of the curves corresponding to different $\xi_i$  in the left panel of Fig.~\ref{fig:evol}. However, as either $\alpha_D$ or $\epsilon$ increases, the temperature ratio given by the attractor solution eventually yields too much energy density in the hidden sector, necessitating a period of reannihilation to obtain the correct late-time relic abundance. Such a trajectory is illustrated by the yellow point (contour) in Fig.~\ref{fig:evol}, left (right). This parameter point demonstrates that at large couplings, a hidden sector that would yield an underabundance of DM in the absence of thermalizing interactions in the hidden sector can develop an overabundance.

Finally, we also show with the orange dashed line in Fig.~\ref{fig:evol} the maximum value of $\epsilon$ for a given $\alpha_D$ for which a traditional freeze-in solution is self-consistent, as given in Eq.~(\ref{eq:alphaepsilonmaxNoScattering}). 
Meanwhile the brown point in the same figure is at the boundary of the region where rapid kinetic equilibration is a good approximation along the freeze-in line ($\alpha_D \gtrsim 10^{-8}$, see Fig.~\ref{fig:K_eq}). 
This leaves a notable portion of the $\alpha_D$-$\epsilon$ plane which can be handled self-consistently in either the non-interacting regime of Sec.~\ref{sec:freezein} or the rapidly thermalizing regime of Sec.~\ref{sec:glaciation}, depending on the presence or absence of a thermalized dark sector.  This should not be a surprise: energetic dark particles produced from the SM plasma are underabundant compared to the thermal number abundance expected for the same $\rho_{\rm HS}$. Thus the rates for self-interactions of these frozen-in particles are small in comparison to the situation where an energetic DM particle with $E_\chi \sim T_{SM}$ scatters off a colder thermal bath of dark particles, even when $\rho_{\rm HS}$ is the same between the two scenarios. This can remain true even if the initial energy density in the dark sector is small, because the dark sector temperature will rapidly approach the attractor solution. Said another way, there are regions of $\epsilon$-$\alpha_D$ space where a minimal dark sector with zero initial abundance may undergo negligible self-scattering, but where a thermal initial population makes the approximation of rapid kinetic equilibration safe.  
This is yet another source of UV sensitivity that goes beyond the dependence on $\xi_i$ demonstrated here.  For instance, the DM produced in the model represented by the brown dot could have very different predictions for its phase space distribution (as well as the number of relic dark mediators), depending on its cosmic history.

Depending on the initial temperature of the hidden sector, the couplings required to achieve the correct DM yield may be considerably smaller than those implied by traditional freeze-in. Indeed, if we drop the requirement of thermal contact with the SM, the kinetic mixing can vanish and the DM can still achieve the observed relic abundance in a decoupled hidden sector. However, if we take some nonzero amount of thermal contact to be a definition of glaciation, we can quantify the UV sensitivity of this scenario in terms of the initial temperature ratio $\xi_i$. As shown in Fig.~\ref{fig:evol} (left), if we fine-tune $\xi_i$ to the critical value $\xi_i^*$ exactly on the overproduced/underproduced boundary of Fig.~\ref{fig:xi_vs_mx}, we end up with the correct relic density by construction even for $\epsilon = \alpha_D = 0$.\footnote{Due to the precise fine-tuning required, the $\xi_i^*$ curve illustrated in Fig.~\ref{fig:evol} saturates at a finite value of $\alpha_D$ due to accumulated rounding error in the numerical solutions to the Boltzmann equations.} For all $\xi_i > \xi_i^*$, decoupled hidden-sector freeze-out with $\epsilon = 0$ is possible, and for all $\xi_i < \xi_i^*$, sufficiently small $\alpha_D$ will permit a traditional freeze-in solution. In this sense, glaciation is UV-sensitive for $\xi_{i} \gtrsim 3 \times 10^{-2}  \left( \mathrm{MeV} / m_{\chi} \right)^{1/3}$. Interestingly, for $\xi_i < \xi_i^*$, there are always points in the $\epsilon$-$\alpha_D$ plane below the traditional freeze-in curve, arising from a period of late-time leak-in supplemented by freeze-in (brown point and yield curve in Fig.~\ref{fig:evol}). To account for this expanded parameter space, we propose that the freeze-in curve should be expanded to a ``glaciation band'' to account for this initial condition sensitivity of freeze-in; we explore the implications of this fact for direct detection experiments below.

\subsection{Implications for direct detection}
\label{sec:dd}

\begin{figure}[!t]
\centering
\includegraphics[width=0.495\textwidth]{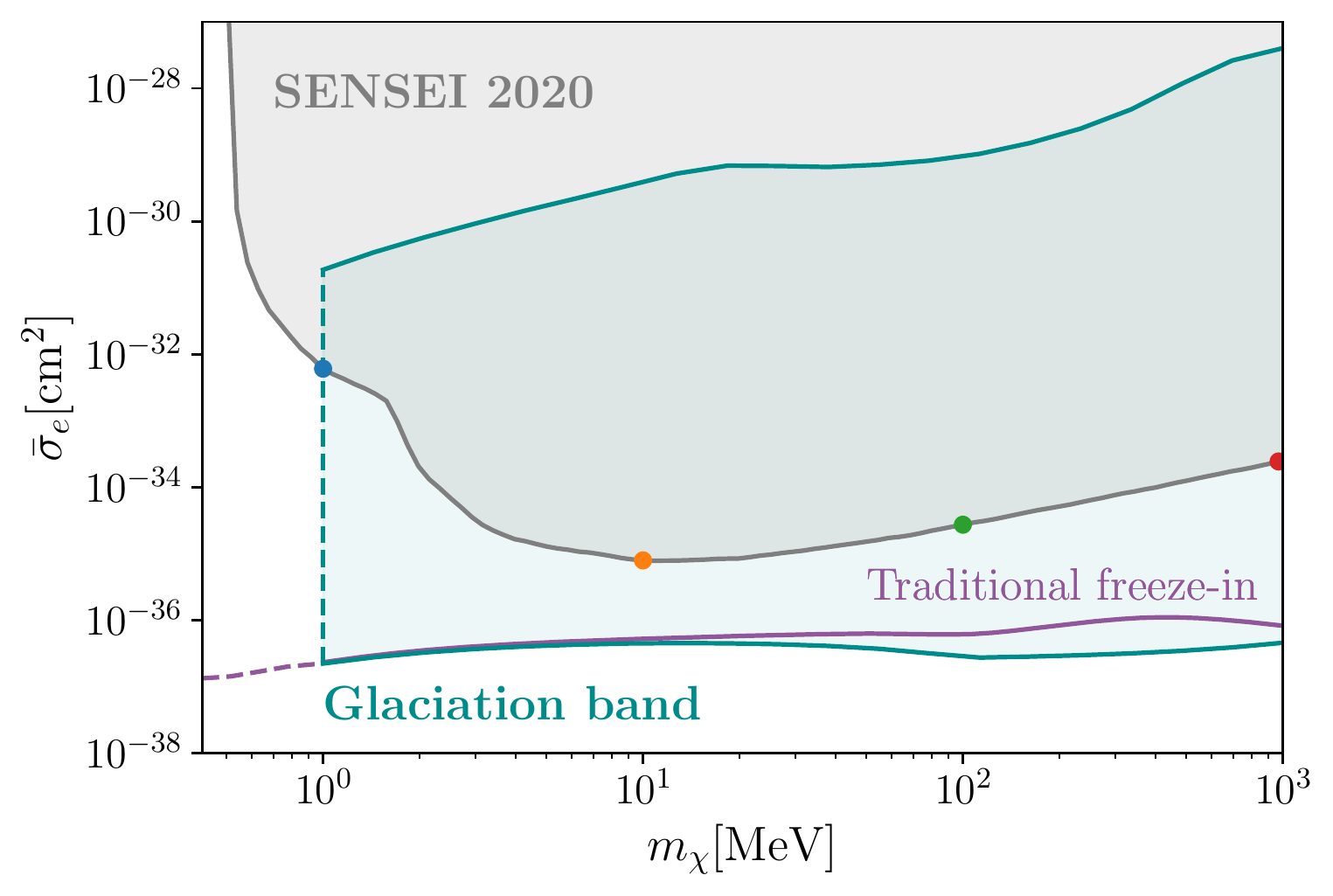}
\includegraphics[width=0.495\textwidth]{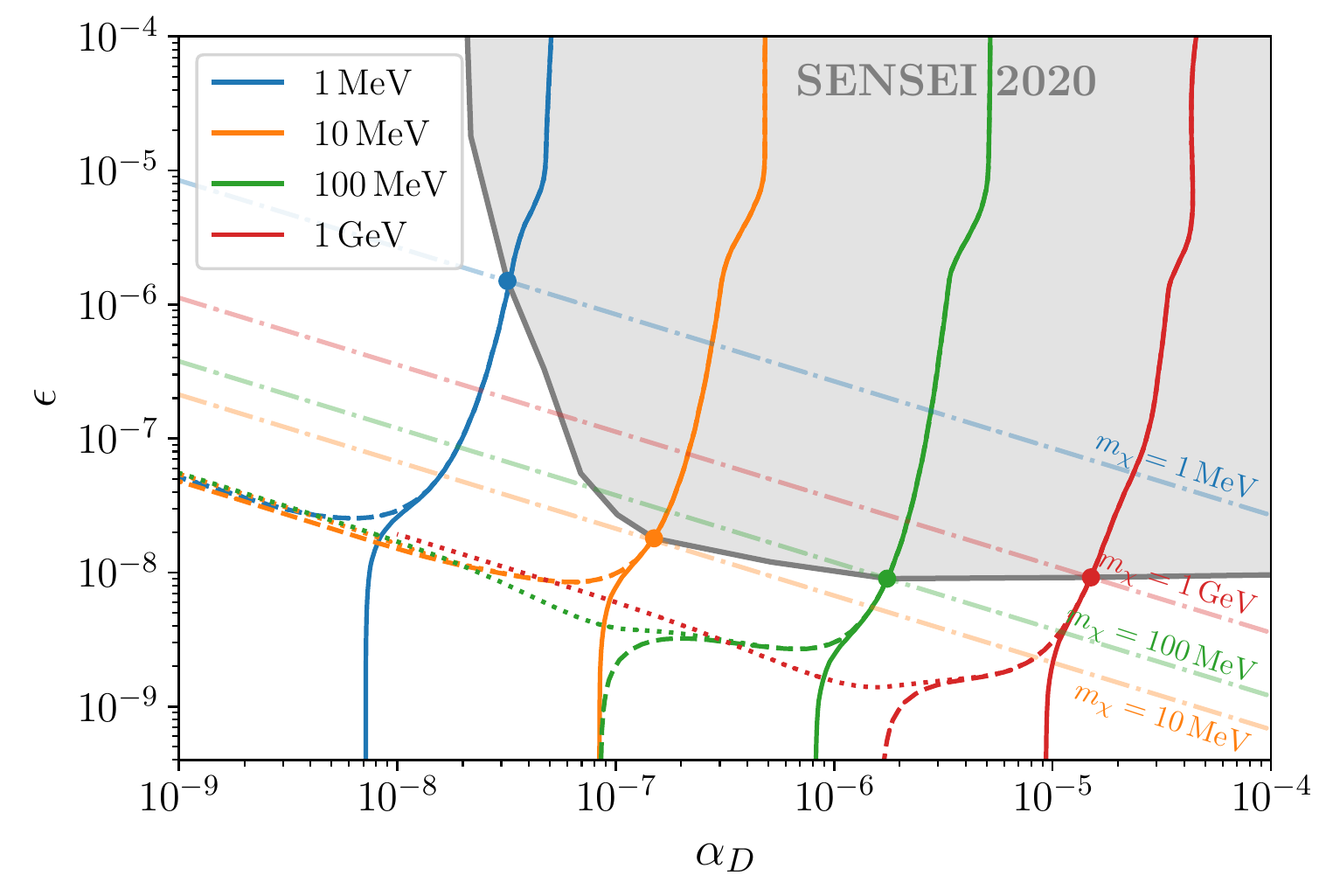}
\caption{\textbf{Left:} The glaciation band on the direct detection parameter space, showing that large regions of UV-insensitive parameter space have already been ruled out by SENSEI. \textbf{Right:} Contours of $\Omega_{\rm DM}/\Omega_{\rm{Obs}} = 1$ in the $\alpha_{D}$-$\epsilon$ plane for different values of DM masses (solid, dashed and dotted lines are for different temperature ratios as in Fig.~\ref{fig:evol}). The intersection of the contours with the direct detection constraints on $\epsilon^2 \alpha_D$ (dash-dotted lines) illustrates the construction of the exclusion region.}
\label{fig:fig_fin}	
\end{figure}

A key feature of the freeze-in scenario is the excellent discovery potential at terrestrial DM-electron scattering experiments, which can take advantage of the low velocity of the DM and the long-range nature of the light $Z_D$ mediator to make up for the small couplings required to match the observed relic abundance. Experimental results are typically expressed in terms of a fiducial DM-electron cross section,
\begin{equation}
    \overline{\sigma}_e \equiv \frac{16\pi \mu_{\chi e}^2 \alpha}{(\alpha m_e)^4} \epsilon^2 \alpha_D \,,
\end{equation}
where $\mu_{\chi e}$ is the DM-electron reduced mass and for simplicity we have assumed $m_{Z_D} \ll \alpha m_e$ in our choice of normalization. Since the dependence of $\overline{\sigma}_e$ on the hidden sector couplings is given by $\epsilon^2 \alpha_D$, we define the \emph{glaciation band} for each DM mass as follows:
\begin{itemize}
    \item Upper boundary: $\epsilon^2 \alpha_D$ equal to the value at the intersection of the $\Omega_\chi/\Omega_{\rm DM} =1$ contour with the thermalization contour (gray dashed in Fig.~\ref{fig:evol}, left).
    \item Lower boundary: $\epsilon^2 \alpha_D$ equal to the minimum value achieved over all contours of $\Omega_\chi/\Omega_{\rm DM} =1$ defined by $\xi_i < \xi_i^*$.
\end{itemize}
By construction, the traditional freeze-in curve is enclosed in the glaciation band. The lower boundary of the glaciation band encompasses the region of parameter space where the relic abundance is dominated by freeze-in processes after accounting for a range of initial conditions. Meanwhile, the upper region of the glaciation band is UV-insensitive, as the attractor solution erases dependence on the initial temperature ratio. We show the glaciation band in Fig.~\ref{fig:fig_fin} (left), along with constraints from SENSEI \cite{Crisler:2018gci,Abramoff:2019dfb,Barak:2020fql} which are the strongest for DM scattering through a light mediator in this region of parameter space. 
We see that direct detection has already ruled out large parts of the glaciation parameter space (see also \cite{Hambye:2018dpi, Evans:2019vxr}). Indeed, this can be visualized in the $\epsilon$-$\alpha_D$ plane as follows. For a given value of $m_\chi$, direct detection sets an upper bound on $\epsilon^2 \alpha_D$, which is a line in the $\epsilon$-$\alpha_D$ plane (dashed lines in Fig.~\ref{fig:fig_fin}, right). The point at which this upper bound intersects the $\Omega_\chi/\Omega_{\rm DM} =1$ contour represents the boundary of the equivalent exclusion region: any larger values of $\epsilon$ and $\alpha_D$ are ruled out, and thus the remainder of the relic density contour for that $m_\chi$ is ruled out. We show the result of this procedure in Fig.~\ref{fig:fig_fin} (right), and see that direct detection constraints already rule out significant portions of the leak-in scenario (independent of $\xi_i$). The projected reach of Oscura \cite{Oscura} will cover the entire glaciation band for $m_\chi > 1 \ {\rm MeV}$; if a positive signal is found at $\overline{\sigma}_e$ below the traditional freeze-in line, that could either indicate a subdominant component of DM, or in the most optimistic case would offer the tantalizing possibility of directly probing the  thermal history of a dark sector with a light mediator.

\section{Conclusions}
\label{sec:conclusions}

Models where DM freezes in through out-of-equilibrium production from the SM have emerged as important targets for developing terrestrial tests of (sub)-GeV-scale dark sectors.  Carefully considering the predictions of freeze-in models is thus vital for understanding the information about the early universe that current and upcoming experiments will provide.

Since DM never reaches thermal equilibrium in freeze-in models, there is necessarily some residual dependence on initial conditions in their predictions.  In the case of ``traditional'' freeze-in, where DM does not interact after its production, this sensitivity is relatively minimal  provided the DM-SM interaction is renormalizable, amounting to a constant and generically small offset of the total DM yield given specific couplings of the DM to the SM.

Another, richer scenario, of high experimental interest, is the case where DM interacts with the SM via a light mediator. Our results here have shown that models where DM freezes in through a kinetically-mixed light mediator can have a much more dramatic dependence on the initial conditions specified for the dark sector than do more traditional freeze-in scenarios.

Using the common and minimal reference model of Dirac fermion dark matter interacting with the SM via a kinetically-mixed dark photon, we have demonstrated a nontrivial dependence of the final DM yield on the initial conditions for the dark photon as well as the dark matter. 
We have shown that it is self-consistent to take the dark sector to be in internal kinetic equilibrium throughout the formation of the DM relic abundance in a large region of interest, and we parameterize the initial conditions for the dark sector in terms of $\xi_i$, the initial ratio of dark to SM temperatures. 

For sufficiently large values of the dark gauge coupling $\alpha_D$ and the kinetic mixing parameter $\epsilon$, the energy injection from the SM is large enough to overwhelm variations in the initial population density, meaning that the DM relic abundance is insensitive to variations in initial conditions.

However, for smaller values of $\alpha_D$ and $\epsilon$, the DM evolution within the hidden sector depends in detail on the initial population.  
In this region, the final DM relic abundance depends on the initial conditions, with different possible outcomes: if the temperature ratio is larger than a critical value $\xi^{*}_{i}$, the evolution of the number density is set by freeze{-}out in the hidden sector, but for smaller initial temperatures the final number density is determined by late freeze{-}in-like processes from the SM. Therefore, the initial population as well as the values  of $\alpha_D$ and $\epsilon$ determine the late-time abundance.  In this region the predicted DM relic abundance exhibits a qualitatively new form of UV sensitivity.

We have pointed out that the freeze-in curve stops being a self-consistent experimental target for sufficiently large values of $\epsilon^2\alpha_D$ and have clarified what happens to hidden sectors with couplings in this regime.  We have shown that a sizeable portion of the resulting ``glaciation band'' is UV-insensitive, in the sense that variations in the initial conditions do not impact the final relic abundance obtained for a given parameter point.  However, for the parameter space near the traditional freeze-in target, predictions for the final relic abundance do depend on the initial population of the hidden sector.  Thus we are able to identify and quantify the residual UV dependence of the freeze-in scenario with light mediators, and clarify its consequences for experiments. We define the bottom of the glaciation band as the smallest SM-DM cross section that gives rise to DM through freeze-in processes from the SM, rather than through hidden sector freeze-out, and provide a simple prescription to compute this quantity.  This glaciation band constitutes a robust and well-motivated target for near future DM-electron direct detection experiments such as Oscura.

Finally, we have provided a simple demonstration that the UV sensitivity of freeze-in with a light mediator goes beyond the dependence on a finite initial temperature.  Since a frozen-in DM particle will scatter much more rapidly off of a cold particle from a pre-existing thermal population than off of another energetic frozen-in DM particle, there are regions of parameter space where {\em both} a non-interacting freeze-in solution and a kinetically-equilibrated glaciation solution can be self-consistent.  In this region the DM phase space distribution will depend on initial conditions even if the DM yield does not.  

In the limit of small $\xi_i$, one may also start to ask whether the hidden sector would have time, in a given cosmological scenario, to approach internal kinetic equilibrium.   The approach to internal thermal equilibrium can take an appreciable amount of time, even for dark sectors containing parametrically light mediators \cite{Garny:2018grs, Forestell:2018dnu, Evans:2019vxr}.  Such questions are particularly acute for the small values of $\alpha_D$ needed to evade constraints on DM self-interactions for sub-MeV DM.  For DM with mass below an MeV, after imposing constraints on DM self-interactions, the approximation of rapid kinetic equilibrium  used here is applicable  for a limited range of relatively large $\xi_i$.  However, the presence of a pre-existing dark sector population, whether equilibrated or not, will generically affect the DM phase space distribution in this mass range as well.  Understanding the impact of scattering in this low-mass region is of particular interest, as the detailed shape of the phase-space distribution of light dark matter can be important for  cosmological observables \cite{Dvorkin:2020xga, DEramo:2020gpr, Decant:2021mhj}.

Determining the evolution of the DM phase-space distribution in the general out-of-equilibrium case requires solving the full Boltzmann hierarchy.  Some work in this direction was recently done in \cite{Du:2021jcj} for a model with a heavy mediator and a constant matrix element.  We expect that this task will be substantially harder for hidden sector with a light mediator, owing to the additional species that needs to be tracked and the need to carefully treat small momentum-transfer scatterings.  However, freeze-in through a kinetically-mixed light dark photon is one of a very small number of cosmologically-viable models for sub-MeV DM, and thus this result is well worth pursuing.

\acknowledgments
YK thanks Sam McDermott and Gordan Krnjaic for helpful conversations in the early stages of this work. The work of NF and JS was supported in part by DOE CAREER grant DE-SC0017840. The work of YK was supported in part by DOE grant DE-SC0015655.

\appendix
\section{Collision terms for species at different temperatures}
\label{sec:collision}

In this appendix, we derive the collision terms for the number density, energy transfer and momentum transfer rates. Unlike the standard case \cite{Gondolo:1990dk, Edsjo:1997bg}, where the species share the same temperature, we generalize the argument and work out the rates for the cases when the initial state particles have different temperatures. Therefore, our general focus will be on processes of the type $1+2 \rightarrow 3+4$, where particles $1$ and $2$ have different bath temperatures $T$ and $\tilde{T}$ respectively, i.e. $T_{1}=T \neq T_{2} =\tilde{T}$. 

We will work under the Maxwell{-}Boltzmann approximation. Therefore, let us first review the relevant thermodynamic equations for particles in thermal equilibrium at temperature $T$ that follow a Maxwell-Boltzmann distribution, namely $f =e^{-\left(E - \mu \right)/T}$, where $E$ is the energy of the particle and $\mu$ its chemical potential.  This leads to the expressions for the number density, energy density, and pressure, given by
\beq
n &= g\dfrac{m^{2} T e^{\mu/T}}{2 \pi^{2}} K_{2} \left( \dfrac{m}{T}\right)  \,, \\
\rho &= g\dfrac{m^{2} T e^{\mu/T}}{2 \pi^{2}} \left[ m K_{1} \left( \dfrac{m}{T}\right) + 3 T K_{2} \left( \dfrac{m}{T}\right)  \right] \,,\\
P &= g\dfrac{m^{2} T^{2} e^{\mu/T}}{2 \pi^{2}} K_{2} \left( \dfrac{m}{T}\right) \,,
\eeq
where $g$ gives the internal degrees of freedom and $K_{i}$ are the modified Bessel functions.

\subsection{Number density}
\label{sec:num_den}

We start with the derivation of the number density collision operator for particle $1$, which reads
\beq
\mathcal{C}_{1\,2 \rightarrow 3\, 4}^{n}(T, \tilde{T}) = -  \int d \Pi_{1} \, d \Pi_{2} \, d \Pi_{3} \, d \Pi_{4}  \abs{\mathcal{M}_{1 \, 2 \rightarrow 3 \, 4 }}^2   (2 \pi)^4  \delta^4(p_{1} + p_{2} - p_{3} - p_{4}) f^{\rm eq}_{1}(T) \,f^{\rm eq}_{2}(\tilde{T})  \, ,
\eeq
where $d \Pi_{i} = d^3p_{i}/2E_{i} \left(2\pi \right)^3$ is the Lorentz-invariant phase space element. For simplicity, here we only consider the collision term governing the forward scattering and neglect the chemical potential, although including these effects is straightforward. The integral over two of the phase space differentials can be written in terms of the cross section $\sigma\left(s \right)$ as 
\beq
 \int d \Pi_{3} \, d \Pi_4  \, \abs{\mathcal{M}_{1 2 \rightarrow 3 4}}^2    (2 \pi)^4  \delta^4(p_{1} + p_{2} - p_{3} - p_4)=  2 g_{1} g_{2} \, \lambda^{1/2}(s,m_{1},m_{2}) \sigma(s)  \ ,
\eeq
where the two-body kinematic function $\lambda (s,m_{1},m_{2})$ is 
\beq
\lambda(s,m_{1},m_{2}) = \lp s - (m_{1}+m_{2})^{2} \rp \lp s - (m_{1} -m_{2})^{2}\rp \,.
\eeq
Then, the collision term can be written as follows
\beq
\mathcal{C}_{1\,2 \rightarrow 3\, 4}^{n}(T, \tilde{T}) = -2 g_{1} g_{2} \int d \Pi_{1} \, d \Pi_{2} \, \lambda^{1/2}(s,m_{1},m_{2}) \, \sigma(s) \,  f^{\rm eq}_{1}(T) \,f^{\rm eq}_{2}(\tilde{T})   \, 
\eeq
and the remaining phase space differentials are, in terms of the lab energies and incident angle,
\beq
d \Pi_{1} \, d \Pi_{2} = \frac{\abs{\vec{p}_{1}} \abs{\vec{p}_{2}} }{32 \pi^{4}} d E_{1} d E_{2} \, d\cos \theta \,.
\eeq 
Here it is convenient to switch the integration variables to
\beq
E_{1} = m_{1} \gamma_{1}  \;\;\;\; E_{2} = m_{2} \gamma_{2}   \;\;\;\; \gamma_{r} = \gamma_{1} \gamma_{2} (1 - \beta_{1}\beta_{2} \cos\theta) = \gamma_{1} \gamma_{2} - \sqrt{\gamma_{1}^{2}-1} \sqrt{\gamma_{2}^{2}-1}  \cos\theta  \,,
\eeq 
where $\gamma$ and $\beta$ are the boost factor and velocity, respectively. The Jacobian for this transformation is
\beq
d \Pi_{1} \, d \Pi_{2} = - \frac{m_{1}^{2} m_{2}^{2}}{32 \pi^{4}} d \gamma_{1} d \gamma_{2} d \gamma_{r}\,,
 \eeq
and the integration limits are
\beq
 a \equiv \gamma_{1}\gamma_{r}  + \gamma_{1}\beta_{1}\gamma_{r}\beta_{r}\leq \,  &\gamma_{2} \leq b \equiv  \gamma_{1}\gamma_{r}  - \gamma_{1}\beta_{1}\gamma_{r}\beta_{r} \\
 &\gamma_{1} \geq  1 \\
 &\gamma_{r} \geq  1   \,.
\eeq
Putting everything together, the collision operator is 
\beq
\mathcal{C}_{1\,2 \rightarrow 3\, 4}^{n}(T, \tilde{T}) = -  \frac{ g_{1} g_{2} m_{1}^{2} m_{2}^{2} }{16 \pi^{4}} \int_{1}^{\infty} d \gamma_{r} \, \sigma(s) \lambda^{1/2}(s,m_{1},m_{2}) \int_{1}^{\infty} d \gamma_{1} \, e^{-\frac{m_{1} \gamma_{1}}{T}} \int_{b}^{a} d \gamma_{2} \, e^{-\frac{m_{2} \gamma_{2}}{\tilde{T}}}  \,.
\eeq
Using the dimensionless variable $x_{i} = m_{i}/T_{i}$ and focusing only on the integration over $\gamma_{1}$ and $\gamma_{2}$, we have
\beq
\int_{1}^{\infty} d \gamma_{1} \, e^{- x_{1} \gamma_{1} } \int_{b}^{a} d \gamma_{2} \,e^{- x_{2} \gamma_{2} }= \frac{1}{x_{2}} \int_{1}^{\infty} d \gamma_{1} \, \lp e^{-x_{1} \gamma_{1} - x_{2} (1 - \beta_{1}\beta_{r}) \gamma_{1} \gamma_{r}} - e^{-x_{1} \gamma_{1} - x_{2} (1 + \beta_{1}\beta_{r}) \gamma_{1} \gamma_{r}} \rp .
\eeq
Then, with the help of the \textit{rapidity} $\gamma_{1} = \cosh (w_{1})$,  $\beta_{1} \gamma_{1} = \sinh (w_{1})$,  $\gamma_{r} = \cosh (w_{r})$ and  $\beta_{r} \gamma_{r} = \sinh (w_{r})$, we can perform the following integration over $w_{1}$ 
 \beq
 \frac{1}{x_{2}}  \int_{0}^{\infty} d w_{1} \, \sinh(w_{1}) \, \lp e^{-x_{1} \cosh (w_{1}) - x_{2}  \cosh (w_{1} - w_{r}) } -  e^{-x_{1} \cosh (w_{1}) - x_{2}  \cosh (w_{1} + w_{r}) }\rp .
 \eeq
Using $\cosh (\theta_{1}) = x_{1}$ and $\cosh (\theta_{2})  = x_{2}$,  the arguments of the exponential can be written as
\barray
\cosh (\theta_{1})  \cosh (w_{1}) + \cosh (\theta_{2}) \cosh (w_{1} \pm w_{r}) = \tilde{s} \cosh(w_{1} \pm \phi ) \, ,
\earray
where 
\beq
\tilde{s} = (x_{1}^{2} + 2 x_{1} x_{2} \gamma_{r} + x_{2}^{2})^{1/2}  \;\;\;\;\; \mathrm{and} \;\;\;\;\; \phi = \sinh ^{-1} \lp \frac{x_{2} \sinh (w_{r})}{\tilde{s}} \rp \,.
\eeq
Shifting the integration variable $t = w_{1} \pm \phi$ we have
\begin{align}
& \frac{1}{x_{2}}  \left( \int_{-\phi}^{\infty} d t \, e^{-  \tilde{s}  \cosh(t)} \sinh(t+\phi)  -  \int_{\phi}^{\infty} d t \, e^{-  \tilde{s}  \cosh(t)} \sinh(t-\phi)  \right) \nonumber \\
= & \frac{1}{x_{2}} \left( \int_{-\infty}^{\infty} d t \, e^{- \tilde{s}  \cosh(t)} \sinh(\phi)\cosh(t)  +  \int_{-\infty}^{\infty} d t \, e^{-  \tilde{s}  \cosh(t)} \sinh(t)\cosh(\phi)  \right) \nonumber \\
=& \frac{2 \sinh(\phi)}{x_{2}}    \int_{0}^{\infty} d t \, e^{- \tilde{s}  \cosh(t)}\cosh(t) \nonumber \\
= &\frac{2 \gamma_{r} \beta_{r} }{\tilde{s}} K_{1} \lp   \tilde{s} \rp \,.
\end{align}
Finally, we find that the collision operator is
\beq
\mathcal{C}_{1\,2 \rightarrow 3\, 4}^{n}(T, \tilde{T}) = -  \frac{ g_{1} g_{2} m_{1}^{2} m_{2}^{2} }{8 \pi^{4}} \int_{1}^{\infty} d \gamma_{r}  \, \frac{ \lambda^{1/2}(s,m_{1},m_{2}) \gamma_{r} \beta_{r}  K_{1}(\tilde{s}) }{\tilde{s}}   \sigma(s).
\eeq
Writing the collision rate in terms of the temperatures and a single integral over $\tilde{s}$, the final result reads
\beq
\mathcal{C}_{1\,2 \rightarrow 3\, 4}^{n}(T, \tilde{T}) =   n_{1\rm eq}(T) n_{2 \rm eq}(\tilde{T})  \langle \sigma v \rangle =  -  \frac{ g_{1} g_{2} \, T^{3} \tilde{T}^{3} }{16 \pi^{4}} \int_{\tilde{s}_{\rm min}}^{\infty} d\tilde{s}  \,  \lambda(\tilde{s}^{2} ,x_{1},x_{2}) K_{1} ( \tilde{s}) \sigma(s)\,,
\label{eq:numberDensityDiff}
\eeq 
where $\tilde{s}_{\rm min}= x_{1} +x_{2}$  and $\sigma(s)$ is evaluated at
\beq
s = \tilde{s}^{2} T \tilde{T} + (T- \tilde{T})(T x_{1}^{2}- \tilde{T} x_{2}^{2})\,
\label{eq:stilde}
\eeq
to perform the integral over $\tilde{s}$. Here $\tilde{s}^2$ plays a role reminiscent of the Mandelstam variable $s$ but now with dependence on the bath temperatures. It is important to note that for elastic scattering processes, the collision term conserves particle number, i.e. $\mathcal{C}_{1\,2 \leftrightarrow 1\, 2}^{n}(T, \tilde{T})=0$. 

\subsection{Energy transfer}
The same calculation can be done for the collision operator describing the energy transfer rate for particle 1 with energy $E=E_{1}$ as
\beq
\mathcal{C}_{1\,2 \rightarrow 3\, 4}^{\rho}(T, \tilde{T}) = -  \int d \Pi_{1} d \Pi_{2} d \Pi_{3} d \Pi_{4}  \abs{\mathcal{M}_{1 \, 2 \rightarrow 3 \, 4 }}^2   E_{1}(2 \pi)^4  \delta^4(p_{1} + p_{2} - p_{3} - p_{4}) f^{\rm eq}_{1}(T) \,f^{\rm eq}_{2}(\tilde{T})  \, .
\eeq
Then after performing a similar calculation as in the number density case, we have that   
\barray
\mathcal{C}_{1\,2 \rightarrow 3\, 4}^{\rho}(T, \tilde{T}) &=&  n_{1\rm eq}(T) n_{2 \rm eq}(\tilde{T})  \langle \sigma v E_{1} \rangle \nonumber \\
&=& -  \frac{ g_{1} g_{2} T^{4} \tilde{T}^{3} }{32 \pi^{4} } \int_{\tilde{s}_{\rm min}}^{\infty} d\tilde{s}  \,  \frac{ \lambda(\tilde{s}^{2} ,x_{1},x_{2})(\tilde{s}^{2}+x_{1}^{2}-x_{2}^{2} )}{\tilde{s}}  \sigma(s) K_{2} ( \tilde{s}) \,.
\label{eq:EnergyTransferDiff}
\earray 
Unlike the number density operator, the energy transfer for elastic scattering processes does not vanish, i.e., $\mathcal{C}_{1\,2 \leftrightarrow 1\, 2}^{\rho}(T, \tilde{T})\neq0$. 

\subsection{Momentum transfer}
\label{app:ptransfer}
Finally, similar to the energy transfer rate, we can define the momentum loss rate of an injected particle with temperature $T$ and momentum $p_1$ through scattering off of a second particle with temperature $\tilde{T}$ and momentum $p_2$ by considering the average of the quantity $\Delta p^{2} = -(p_{1}-p_{3})^{2}=2\vec{p}_{1}^{\,2}(1-\cos\theta)$, where  $p_3$ the momentum of the injected particle after the collision and in the second equality $\vec p_1$ and the scattering angle are given in the center-of-mass frame. This expression for the momentum transfer-squared is just the Mandelstam variable $-t$ in the center of mass frame. Then, the collision operator describing the momentum loss rate can be written as
\barray
\mathcal{C}_{1\,2 \rightarrow 3\, 4}^{p}(T, \tilde{T}) &=& n_{1\rm eq}(T) n_{2 \rm eq}(\tilde{T})  \langle \sigma v  \Delta p^{2} \rangle \\
&=& -  \frac{ g_{1} g_{2} T^{5} \tilde{T}^{3} }{32 \pi^{4} } \int_{\tilde{s}_{\rm min}}^{\infty} d\tilde{s}  \,  \frac{ \lambda(\tilde{s}^{2} ,x_{1},x_{2})}{\tilde{s}^{2}}   \sigma_{T}(s) \lp \lambda(\tilde{s}^{2} ,x_{1},x_{2}) K_{3}(\tilde{s})  + 4 \tilde{s} x_{1}^{2} K_{1}(\tilde{s})  \rp \nonumber \,,
\earray 
where $ \sigma_{T}$ is the transfer cross section defined by $\sigma_{T} = \int \frac{d \sigma}{d \Omega}  (1-\cos\theta) d \Omega$ \cite{Buckley:2009in,Feng:2009hw}. Thus, using the previous results, the thermally-averaged momentum loss rate can be obtained by 
\beq
 \Gamma_{p \, \mathrm{loss}} \approx \langle \frac{d \Delta p^{2}}{dt} \rangle  \frac{1}{ \langle p_{1}^{2} \rangle } = \frac{n_{2 \rm eq}(\tilde{T}) \langle \sigma v  \Delta p^{2} \rangle }{\langle p_{1}^{2} \rangle}\,,
\eeq
where $\langle p_{1}^{2} \rangle$, the average momentum-squared of an injected DM particle, is
\beq
\langle p_{1}^{2} \rangle= \frac{\int d^{3} p \, p^{2} f(p)}{\int d^{3} p  \,f(p)}=3 m_{1} T \lp \frac{K_{3} \lp m_{1}/T \rp}{K_{2} \lp m_{1}/T \rp}\rp \,.
\eeq


\begin{figure}[!t]
\center
\includegraphics[width=0.7\textwidth]{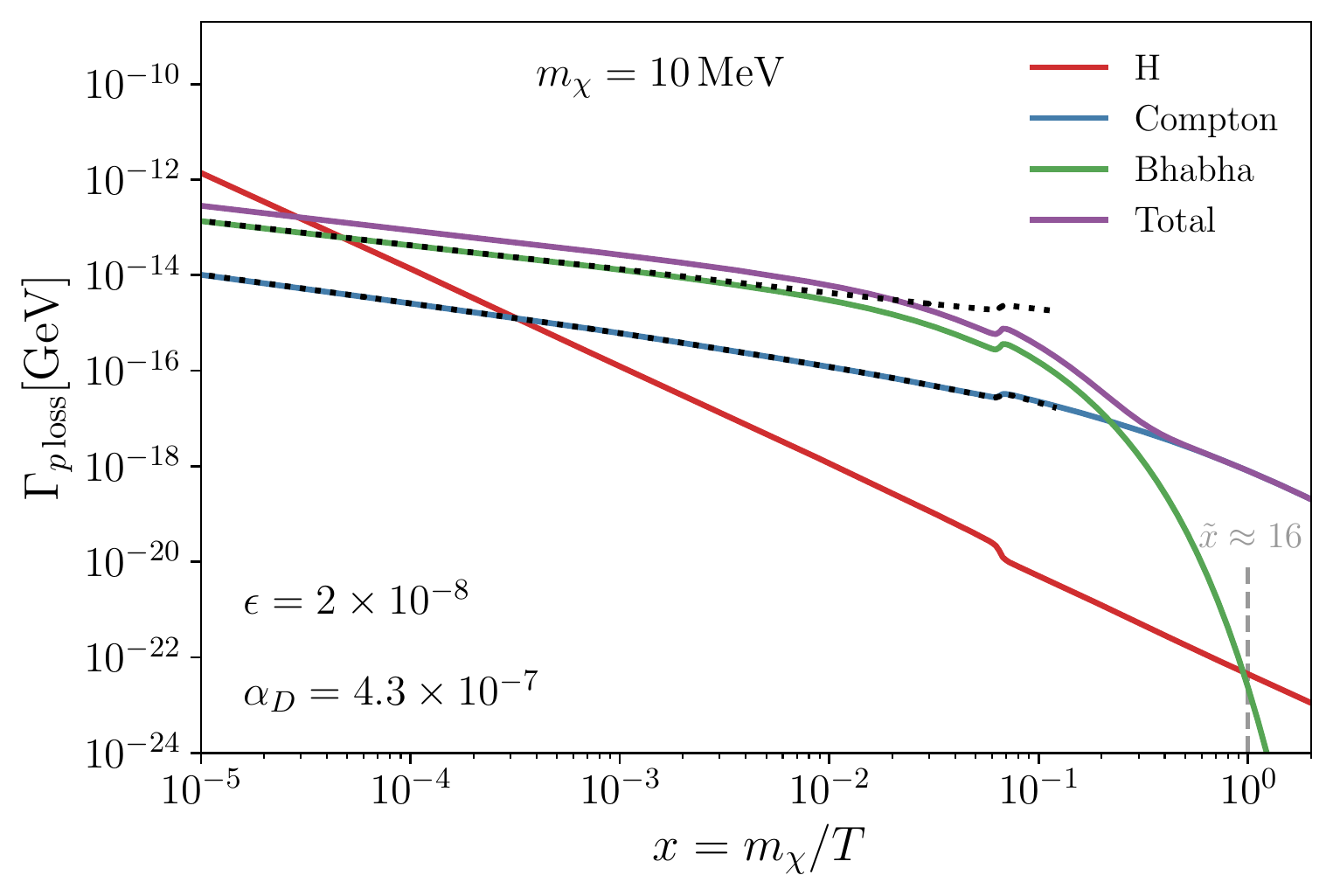}
\caption{Momentum loss rate due to various scattering processes compared to the Hubble rate. Here, we are accounting for the non-adiabatic evolution of the hidden sector temperature in evaluating the rates. The parameters used are the same as the yellow point in Fig. \ref{fig:evol}, with $\xi_{i}=10^{-3}$ and $m_{Z_D} = 10^{-15} \ \mathrm{ eV}$. The Bhabha and M\o ller rates are identical so for clarity we only show one curve. 
Finally, the black dotted lines are the semi{-}analytical
solutions for the rates for $T \gg m_{\chi}$ and the gray dashed line is the value of $\tilde{x}$ at $x=1$ to illustrate the Boltzmann suppression for the Bhabha and M\o ller rates.}

\label{fig:Scattering}
\end{figure}

\subsection{Rapid kinetic equilibration}
\label{app:IKE}

As explained in the text, we can now estimate when rapid kinetic equilibrium holds by requiring $ \Gamma_{p \, \mathrm{loss}} \gtrsim H$. Our goal here is to verify that our choice of initial conditions for solving the Boltzmann equations is robust: if rapid kinetic equilibrium is obtained at some point while the DM is relativistic, it is maintained throughout all of the evolution of the DM number density. We are interested in the elastic scattering processes --  namely Compton, Bhabha, and M\o ller -- explicit cross-sections for which are given in Appendix~\ref{app:CS}. Fig.~\ref{fig:Scattering} shows the momentum loss rate from each of these processes separately as well as the total, together with the Hubble parameter. For this example, we have chosen parameter values corresponding to the yellow point in Fig.~\ref{fig:evol} for which the value of $\epsilon^{2} \alpha_{D}$ is small, meaning the energy transfer is small too and $\xi_{i}=10^{-3}$. In this case, as can been seen in the right panel of Fig. \ref{fig:evol}, the hidden temperature starts evolving non{-}adiabatically on the attractor solution.  For this parameter point, over the range of hidden sector temperatures for which we solve the Boltzmann equations (starting at $x_{i}=10^{-2} \xi_{i}$), the total $\Gamma_{p \, \mathrm{loss}}$ is always larger than the Hubble rate. Therefore, there is always self-consistency when the equations are solved numerically. To gain some intuition, let us consider the limiting case of $T \gg m_{\chi}$, where we can obtain an approximate analytic expression to the momentum loss rate due to Bhabha scattering, as 
\beq
\Gamma^{\mathrm{Bhabha}}_{p \, \mathrm{loss}} \approx  \frac{ \alpha_{D}^{2}  T \xi^{2}}{6 \pi}  \left( 47 - 48\gamma_{E}+ 24 \log\left(\xi  \frac{T^{2}}{m^{2}_{Z_{D}}} \right)\right) ,
\eeq
and for Compton, as
\beq
\Gamma^{\mathrm{Compton}}_{p \, \mathrm{loss}} \approx  \frac{2 \alpha_{D}^{2}  T \xi^{2}}{\pi}  \left(1 - 2\gamma_{E} + \log\left(4\xi  \frac{T^{2}}{m^{2}_{\chi}} \right)\right) \, ,
\eeq
where $\gamma_{E}$ is Euler’s constant. If the hidden sector evolves adiabatically, $\xi$ is a constant, in contrast to the non{-}adiabatic case, where the attractor solution is well{-}approximated by Eq.~(\ref{eq:xiLI}). While in the former case the rates scale as $T$, as generically expected, in the latter case the scaling is $T^{1/2}$. As can be seen in Fig.~\ref{fig:Scattering} (black dotted lines), the semi{-}analytical estimates track the numerical solution perfectly. Thus, the approximate minimum value $x_{\mathrm{min}}$ that ensures kinetic equilibrium will satisfy $\Gamma^{\mathrm{Bhabha}}_{p \, \mathrm{loss}} \left( x_{\mathrm{min}} \right) \approx H\left( x_{\mathrm{min}} \right)$, which gives
\beq
x_{\mathrm{min}} \approx 2 \times 10^{-5} \left(\frac{4.3 \times 10^{-7}}{\alpha_D} \right)^{5/3}  \left( \dfrac{2\times 10^{-8}}{\epsilon}\right)^{2/3} \left( \frac{m_{\chi}}{10 \MeV} \right)\,.
\eeq
Moreover, we also need to ensure that kinetic equilibrium is maintained until the final DM number density has been achieved. For the parameter space considered here, we find that the Compton rate always preserves the kinetic equilibrium conditions for late times as shown in Fig. \ref{fig:Scattering}. The reason is that the Compton rate is not Boltzmann-suppressed, and furthermore at late times when the energy injection from the SM is negligible, the Compton rate has the same temperature scaling as the Hubble parameter, $H \propto T^{2}$. Therefore, once Hubble crosses the Compton rate from above, Compton dominates for all late times and kinetic equilibrium is maintained.

\subsection{Number and energy density collision terms for $\tilde{T}=T$}
\label{app:collision_terms}
We can compute the thermal average for the annihilation cross section and the energy transfer rate for $f\bar{f}\rightarrow \chi \bar{\chi}$ using Eqs.~(\ref{eq:numberDensityDiff}) and (\ref{eq:EnergyTransferDiff}), 
and demonstrate that our two-temperature result gives the correct answer in the limit that the two temperatures are equal. Here the SM fermions are in equilibrium with the SM thermal bath, i.e.~$\tilde{T}=T$, and thus Eq.~(\ref{eq:stilde}) takes the simple form $s=\tilde{s}^{2}T^{2}$, giving $\tilde{s}=\sqrt{s}/T$ and  $d\tilde{s} = ds/(2T\sqrt{s})$. Inserting these factors into Eq.~(\ref{eq:numberDensityDiff}), the collision term for annihilation is
\beq
\mathcal{C}_{f\bar{f}\rightarrow \chi \bar{\chi}}^{n}(T) =   \langle \sigma v \rangle_{\rm fi}\, n_{f}^{2}(T) =  \frac{ g_{f} g_{\bar{f}} \, T }{32 \pi^{4}} \int_{s_{\rm min}}^{\infty} ds  \, \sqrt{s}(s-4m_{f}^{2}) \, \sigma_{f\bar{f}\rightarrow \chi \bar{\chi}}(s)\, K_{1} (\sqrt{s}/T),
\label{eq:numberDensityffxx}
\eeq 
where $s_{\rm min} = \mathrm{max} \{ 4m^{2}_{f},4m^{2}_{\chi}\}$. This recovers the well-known results from Ref.~\cite{Gondolo:1990dk}.
Similarly, using Eq.~(\ref{eq:EnergyTransferDiff}) the energy transfer rate reads
\barray
\mathcal{C}_{f\bar{f}\rightarrow \chi \bar{\chi}}^{\rho}(T) &=& \langle \sigma v E \rangle_{\rm fi}\, n_{f}^{2}(T)  \nonumber \\
&=&  \frac{ g_{f} g_{\bar{f}} T }{32 \pi^{4} } \int_{s_{\rm min}}^{\infty} ds  \,  s (s-4m_{f}^{2}) \sigma_{f\bar{f}\rightarrow \chi \bar{\chi}}(s)\, K_{2} (\sqrt{s}/T) \,,
\label{eq:EnergyTransferffxx}
\earray 
where $E=E_{1}+E_{2}$ and we have used $\langle \sigma v E \rangle_{\rm fi}= 2 \langle \sigma v E_{1} \rangle_{\rm fi}$. 
Finally, we provide explicit formulae for the number and energy density rates for $Z$ decays into DM, 
\beq
&\mathcal{C}_{Z\rightarrow \chi \bar{\chi}}^{n}(T) =  \langle \Gamma \rangle_{Z} n_{Z}(T) =  \frac{g_{Z}m^{2}_{Z}T}{2 \pi^{2}} \Gamma_{Z \rightarrow \chi \bar\chi} K_{1}\lp m_{Z}/T\rp \,,\\
&\mathcal{C}_{Z\rightarrow \chi \bar{\chi}}^{\rho}(T) = \langle \Gamma  E_{Z}\rangle_{Z} n_{Z}(T) =  \frac{g_{Z}m^{3}_{Z}T}{2 \pi^{2}} \Gamma_{Z \rightarrow \chi \bar\chi} K_{2}\lp m_{Z}/T\rp\,,
 \label{eq:DecayTransferffxx}
\eeq
where $g_{Z}=3$ gives the degrees of freedom of the $Z$ boson.
\section{Cross sections}
\label{app:CS}
For reference, we present all of the $2 \to 2$ cross section and decay formulas we require in our Boltzmann equations. All cross sections here are summed, rather than averaged, over the final and initial states.

\subsection*{Decay of $Z$ to DM, $Z \rightarrow \chi \bar\chi$}
The total decay width is
 \begin{align}
\Gamma_{Z \rightarrow \chi \bar\chi} = \frac{ (\epsilon g_{D}  \tan\theta_W)^{2} m_{Z}}{12 \pi}   \lp 1+ \frac{2 m^{2}_{\chi}}{m^{2}_{Z}} \rp \sqrt{1- \frac{4 m^{2}}{m^{2}_{Z}}} \,,
\end{align}
where $\theta_W$ is the weak mixing angle. 
\subsection*{SM fermion annihilations to DM only through the dark photon, $f \bar f \rightarrow \chi \bar\chi$}
\beq
\sigma_{f \bar f \rightarrow \chi \bar\chi } (s) = \frac{(\epsilon e Q_{f} g_{D})^2}{3 \pi}   \lp  \frac{\sqrt{s-4 m_{\chi}}}{\sqrt{s-4 m_{f}}} \frac{ (s+2m^{2}_{\chi}) (s+2m^{2}_{f}) }{s^{3}} \rp.
\eeq

\subsection*{SM fermion annihilations to DM with $Z_{D}-Z$ contribution, $f \bar f \rightarrow \chi \bar\chi$}
Here we show the full annihilation cross section including the $Z$ boson contribution,
\beq
 \sigma_{f \bar f \rightarrow \chi \bar\chi } (s)=&\frac{\sqrt{s-4 m_{\chi}^{2}}}{\pi s \sqrt{s-4 m_{f}^{2}}} \lp \frac{(\epsilon e Q_{f} g_{D})^2}{3} \lp \frac{ (s+2m^{2}_{\chi}) (s+2m^{2}_{f}) }{s^{2}} \rp \nonumber \right. \\
-& \frac{\epsilon^{2} e Q_{f} g^{2}_{D} g_{Z}  \tan\theta_W  C_{V} }{ 3}   \lp \frac{(s+2m^{2}_{\chi}) (s+2m^{2}_{f}) (s-m^{2}_{Z})}{s \lp (s-m^{2}_{Z})^{2} +m^{2}_{Z} \Gamma^{2}_{Z}\rp} \rp  \nonumber \\
+& \left.\frac{(\epsilon g_{Z} g_{D}  \tan\theta_W)^2 }{12}   \lp \frac{(s+2m^{2}_{\chi})\lp C^{2}_{V}(s + 2m^{2}_{f} )  + C^{2}_{A}(s-4m^{2}_{f}) \rp  }{ (s-m^{2}_{Z})^{2} +m^{2}_{Z} \Gamma^{2}_{Z}} \rp \rp\,,
\eeq
where the vector and axial couplings for a fermion $f$ are  $C_{V} = T_{f}^{3} - 2 Q_{f} \sin^{2}{\theta_W}$,  $C_{A}=T_{f}^{3}$, $g_{Z}= \frac{e}{\cos\theta_W \sin\theta_W}$ and $\Gamma_{Z}$ is the decay width of the $Z$ boson. 

\subsection*{DM annihilations to dark photons, $\chi \bar\chi  \rightarrow  Z_{D}  Z_{D}$}
\beq
\sigma_{\chi \bar\chi  \rightarrow  Z_{D}  Z_{D}} (s) = \frac{ g_{D}^4}{4 \pi \, s}   \lp  \frac{2 \lp s^{2} +4 s m_{\chi}^{2} -8 m_{\chi}^{4} \rp}{s(s-4m_{\chi}^{2})}\tanh^{-1} \lp \sqrt{\frac{s-4 m_{\chi}^{2}}{s}}\rp   - \frac{ (s+4m^{2}_{\chi}) }{\sqrt{s (s- 4 m_{\chi}^{2})}} \rp.
\eeq
\subsection*{Dark Compton scattering, $\chi Z_{D}   \rightarrow   \chi  Z_{D}$}
\begin{align}
\sigma_{\chi Z_{D}   \rightarrow   \chi  Z_{D}} (s) = \frac{ g_{D}^4}{4 \pi \, s^2 \lp s-m_{\chi}^{2}\rp^{3}}  & \lp  2 s^{2} \lp s^{2} - 6 s m_{\chi}^{2} -3 m_{\chi}^{4} \rp \log\lp \frac{s}{m_{\chi}^{2}}\rp \right. \nonumber \\
 & + \lp s- m_{\chi}^{2} \rp \lp s^{3} + 15 s^{2} m_{\chi}^{2} - s m_{\chi}^{4} + m_{\chi}^{6}\rp \bigg ).
\end{align}

\subsection*{Dark Bhabha scattering, $\chi \bar\chi   \rightarrow  \chi \bar\chi $}
The Bhabha scattering cross section in the limit of $m_{Z_{D}} \ll m_{\chi}$  is, up to $\mathcal{O}(m^{2}_{Z_{D}}/s)$,
 \beq
\sigma_{\chi \bar\chi   \rightarrow  \chi \bar\chi } (s) \approx & \frac{ g_{D}^4 }{ \pi  s \lp s-4 m_{\chi}^{2}\rp}\left(  \frac{   \lp s-2 m_{\chi}^{2}\rp^{2}}{ m_{Z_{D}}^{2}} +  \right.\\
& \left. \frac{ 7 s^{4} -46 s^{3} m_{\chi}^{2} + 24 s^{2} m_{\chi}^{4} + 128s  m_{\chi}^{6} +64 m^{8}_{\chi} - 6 s \lp s^{2}-2 m_{\chi}^{4} \rp \lp s-4m_{\chi}^{2} \rp  \log \lp \frac{s-4 m_{\chi}^{2}}{m_{Z_{D}}^{2}} \rp }{3 \, s^{2} \lp s-4 m_{\chi}^{2}\rp} \right).
\eeq
\subsection*{Dark M\o{}ller scattering, $\chi \chi   \rightarrow  \chi \chi $}
The M\o{}ller scattering cross section in the limit of $m_{Z_{D}} \ll m_{\chi}$  is, up to $\mathcal{O}(m^{2}_{Z_{D}}/s)$,
\beq
\sigma_{\chi \chi   \rightarrow  \chi \chi } (s) \approx \frac{ \, g_{D}^4 }{ \pi s \lp s-4 m_{\chi}^{2}\rp} \left( \frac{  \lp s-2 m_{\chi}^{2}\rp^{2}}{ m_{Z_{D}}^{2} }  + \frac{ s^2 - 8 s m^{2}_{\chi} + 8m^{4}_{\chi} - 8 m^{2}_{\chi}\lp s-3 m_{\chi}^{2} \rp \log \lp \frac{s-4 m_{\chi}^{2}}{m_{Z_{D}}^{2}} \rp }{2 \lp s-4 m_{\chi}^{2}\rp } \right).
\eeq
Notice Bhabha and M\o{}ller scattering cross-sections agree at leading order in $m^{2}_{Z_{D}}/s$.
\bibliographystyle{JHEP}
\bibliography{references}

\end{document}